\documentclass{aastex62}

\newcommand{\ztf}{ZTF\,J2130+4420}
\newcommand{\kms}{\ensuremath{{\rm km}\,{\rm s}^{-1}}}
\newcommand{\degree}{$^{\circ}$ }
\newcommand{\msol}{M$_\odot$}
\newcommand{\rsol}{R$_\odot$}
\newcommand{\teff}{$T_{\rm eff}$}
\newcommand{\logg}{$\log(g)$}
\newcommand{\porb}{$P_{\rm orb}$}
\newcommand{\vrot}{$v_{\rm rot}\sin{i}$}
\usepackage{multirow}
\usepackage{xcolor}
\usepackage{amssymb}

\graphicspath{{./}{figures/}}

\received{January 1, 2018}
\revised{January 7, 2018}
\accepted{\today}
\submitjournal{ApJ}

\shorttitle{An ultracompact accreting sdO binary}
\shortauthors{Kupfer et al.}
\begin{document}

\title{The first ultracompact Roche lobe-filling hot subdwarf binary}

\correspondingauthor{Thomas Kupfer}
\email{tkupfer@ucsb.edu}

\author[0000-0002-6540-1484]{Thomas Kupfer}
\affiliation{Kavli Institute for Theoretical Physics, University of California, Santa Barbara, CA 93106, USA}

\author[0000-0002-4791-6724]{Evan B. Bauer}
\affiliation{Kavli Institute for Theoretical Physics, University of California, Santa Barbara, CA 93106, USA}

\author[0000-0002-2498-7589]{Thomas R. Marsh}
\affiliation{Department of Physics, University of Warwick, Coventry CV4 7AL, UK}

\author[0000-0002-2626-2872]{Jan van~Roestel}
\affiliation{Division of Physics, Mathematics and Astronomy, California Institute of Technology, Pasadena, CA 91125, USA}

\author[0000-0001-8018-5348]{Eric C. Bellm}
\affiliation{DIRAC Institute, Department of Astronomy, University of Washington, 3910 15th Avenue NE, Seattle, WA 98195, USA}

\author[0000-0002-7226-836X]{Kevin B. Burdge}
\affiliation{Division of Physics, Mathematics and Astronomy, California Institute of Technology, Pasadena, CA 91125, USA}

\author[0000-0002-8262-2924]{Michael W. Coughlin}
\affiliation{Division of Physics, Mathematics and Astronomy, California Institute of Technology, Pasadena, CA 91125, USA}

\author{Jim Fuller}
\affiliation{Division of Physics, Mathematics and Astronomy, California Institute of Technology, Pasadena, CA 91125, USA}

\author[0000-0001-5941-2286]{JJ Hermes}
\affiliation{Department of Astronomy, Boston University, 725 Commonwealth Ave., Boston, MA 02215, USA}

\author{Lars Bildsten}
\affiliation{Kavli Institute for Theoretical Physics, University of California, Santa Barbara, CA 93106, USA}
\affiliation{Department of Physics, University of California, Santa Barbara, CA 93106, USA}

\author[0000-0001-5390-8563]{Shrinivas R. Kulkarni}
\affiliation{Division of Physics, Mathematics and Astronomy, California Institute of Technology, Pasadena, CA 91125, USA}

\author{Thomas A. Prince}
\affiliation{Division of Physics, Mathematics and Astronomy, California Institute of Technology, Pasadena, CA 91125, USA}

\author[0000-0003-4373-7777]{Paula Szkody}
\affiliation{University of Washington, Department of Astronomy, Box 351580, Seattle, WA 98195, USA}

\author[0000-0003-4236-9642]{Vik S. Dhillon}
\affiliation{Department of Physics \& Astronomy, University of Sheffield, Sheffield S3 7RH, UK}
\affiliation{Instituto de Astrofísica de Canarias, Via Lactea s/n, La Laguna, E-38205 Tenerife, Spain}

\author[0000-0001-7809-1457]{Gabriel Murawski}
\affiliation{Gabriel Murawski Private Observatory (SOTES), Poland}

\author{Rick Burruss}
\affiliation{Caltech Optical Observatories, California Institute of Technology, Pasadena, CA 91125, USA}

\author{Richard Dekany}
\affiliation{Caltech Optical Observatories, California Institute of Technology, Pasadena, CA 91125, USA}

\author{Alex Delacroix}
\affiliation{Caltech Optical Observatories, California Institute of Technology, Pasadena, CA 91125, USA}

\author{Andrew J. Drake}
\affiliation{Division of Physics, Mathematics and Astronomy, California Institute of Technology, Pasadena, CA 91125, USA}

\author[0000-0001-5060-8733]{Dmitry A. Duev}
\affiliation{Division of Physics, Mathematics and Astronomy, California Institute of Technology, Pasadena, CA 91125, USA}

\author{Michael Feeney}
\affiliation{Caltech Optical Observatories, California Institute of Technology, Pasadena, CA 91125, USA}

\author[0000-0002-3168-0139]{Matthew J. Graham}
\affiliation{Division of Physics, Mathematics and Astronomy, California Institute of Technology, Pasadena, CA 91125, USA}

\author[0000-0001-6295-2881]{David~L.\ Kaplan}
\affiliation{Center for Gravitation, Cosmology and Astrophysics, Department of Physics, University of Wisconsin--Milwaukee, P.O.\ Box 413, Milwaukee, WI 53201, USA}

\author[0000-0003-2451-5482]{Russ R. Laher}
\affiliation{IPAC, California Institute of Technology, 1200 E. California Blvd, Pasadena, CA 91125, USA}

\author[0000-0001-7221-855X]{S. P. Littlefair}
\affiliation{Department of Physics \& Astronomy, University of Sheffield, Sheffield S3 7RH, UK}

\author[0000-0002-8532-9395]{Frank J. Masci}
\affiliation{IPAC, California Institute of Technology, 1200 E. California Blvd, Pasadena, CA 91125, USA}

\author{Reed Riddle}
\affiliation{Caltech Optical Observatories, California Institute of Technology, Pasadena, CA 91125, USA}   

\author[0000-0001-7648-4142]{Ben Rusholme}
\affiliation{IPAC, California Institute of Technology, 1200 E. California Blvd, Pasadena, CA 91125, USA}

\author{Eugene Serabyn}
\affiliation{Jet Propulsion Laboratory, California Institute of Technology, Pasadena, CA 91109, USA}

\author{Roger M. Smith}
\affiliation{Caltech Optical Observatories, California Institute of Technology, Pasadena, CA 91125, USA}

\author[0000-0003-4401-0430]{David L. Shupe}
\affiliation{IPAC, California Institute of Technology, 1200 E. California Blvd, Pasadena, CA 91125, USA}
             
\author[0000-0001-6753-1488]{Maayane T. Soumagnac}
\affiliation{Lawrence Berkeley National Laboratory, 1 Cyclotron Road, Berkeley, CA 94720, USA}
\affiliation{Department of Particle Physics and Astrophysics, Weizmann Institute of Science, Rehovot 76100, Israel}



\begin{abstract}

We report the discovery of the first short period binary in which a hot subdwarf star (sdOB) fills its Roche lobe and started mass transfer to its companion. The object was discovered as part of a dedicated high-cadence survey of the Galactic Plane named the Zwicky Transient Facility and exhibits a period of $P=39.3401(1)$\,min, making it the most compact hot subdwarf binary currently known. Spectroscopic observations are consistent with an intermediate He-sdOB star with an effective temperature of \teff$=42,400\pm300$\,K and a surface gravity of \logg$=5.77\pm0.05$. A high-signal-to noise GTC+HiPERCAM light curve is dominated by the ellipsoidal deformation of the sdOB star and an eclipse of the sdOB by an accretion disk. We infer a low-mass hot subdwarf donor with a mass $M_{\rm sdOB}=0.337\pm0.015$\,\msol\, and a white dwarf accretor with a mass $M_{\rm WD}=0.545\pm0.020$\,\msol.

Theoretical binary modeling indicates the hot subdwarf formed during a common envelope phase when a $2.5-2.8$\,\msol\, star lost its envelope when crossing the Hertzsprung Gap. To match its current \porb, \teff, \logg, and masses, we estimate a post-common envelope period of \porb$\approx150$\,min, and find the sdOB star is currently undergoing hydrogen shell burning. We estimate that the hot subdwarf will become a white dwarf with a thick helium layer of $\approx0.1$\,\msol\, and will merge with its carbon/oxygen white dwarf companion after $\approx17$\,Myr and presumably explode as a thermonuclear supernova or form an R\,CrB star. 

%To put constraints on the structure and evolutionary history of the system, we calculated evolutionary models with Modules for Experiments in Stellar Astrophysics (\texttt{MESA}). The hot subdwarf formed during a common envelope phase when a $2.5-2.8$\,\msol\,main sequence star lost its envelope when crossing the Hertzsprung Gap. To match its current \porb, \teff, \logg, and masses, we find that the object left the common envelope at \porb$=148$\,min and reached contact at \porb$=40.5$\,min after helium core burning and residual hydrogen shell burning expanded the radius of the hot subdwarf beyond its Roche radius. After hydrogen shell burning is finished, the hot subdwarf will become a white dwarf with a thick helium layer of $\approx0.15$\,\msol\, which will merge with its carbon/oxygen white dwarf companion after $\approx17$\,Myr and either explode as a thermonuclear supernova or form an R\,CrB star.  

\end{abstract}



\section{Introduction} \label{sec:intro}
Hot subdwarf B/O (sdB/O) stars are stars with spectral type B or O but less luminous than main sequence stars with the same spectral type. Most are thought to be compact He-burning stars with masses around $0.5$\,\msol\, and with thin hydrogen envelopes \citep{heb86, heb09, heb16}. It has been shown that a large number of sdB stars are in close orbits with orbital periods of  \porb$<10$\,days \citep{nap04a,max01}. The most compact systems have orbital periods of $\lesssim 1$\,hour (e.g. \citealt{ven12,gei13,kup17,kup17a}). A possible mechanism that may form such tight binaries is orbital shrinkage through a common envelope phase, followed by the loss of angular momentum due to the radiation of gravitational waves \citep{han02,han03,nel10a}.

Hot subdwarf binaries with white dwarf (WD) companions that exit the common envelope phase at \porb$\lesssim$2\,hours will overflow their Roche lobes while the sdB is still burning helium. Due to the emission of gravitational waves, the orbit of the binary will shrink until the sdB/O fills its Roche lobe at a period of $\approx20$-$40$\,min, depending on the evolutionary stage of the hot subdwarf (e.g. \citealt{sav86,tut89,tut90,ibe91,yun08,pie14,bro15}). However, no  hot subdwarf in a tight accreting binary that fills its Roche lobe has been found so far. 

%\trmc{Is it worth saying here that to date no such system (i.e. with a Roche lobe filling hot subdwarf) has been found?}

The known population of sdB binaries consist mostly of systems with orbital periods too large to initiate accretion before the sdB evolves into a WD \citep{kup15a}. Currently only three systems with a WD companion are known to have \porb$<2$\,hours \citep{ven12,gei13,kup17,kup17a}. 

\citet{kup17a} discovered the most compact binary consisting of a hot subdwarf star with a massive WD companion as part of the OmegaWhite survey \citep{mac15,tom16,mac17}. OW\,J074106.0--294811.0 has a period of only 44\,min. However, the analysis revealed that the hot subdwarf has a mass of only $0.23\pm0.12$\msol\, and is inconsistent with a canonical-mass helium burning sdB star; instead, it is fully consistent with a helium-core WD with a mass of $0.32$\,\msol. If the hot subdwarf star is a helium WD, the system will start accretion at an orbital period of $\approx$5\,min. Depending on the spin-orbit synchronization timescale, the object will either merge and form an R\,CrB star or end up as a stable accreting AM\,CVn type system with a He WD donor \citep{kup17a}. 

The most compact known sdB binary where the sdB is still undergoing helium-core burning is CD--30$^{\circ}$11223. The binary has an orbital period \porb=70.5\,min and a high mass WD companion ($M_{\rm WD}\approx0.75$\,\msol; \citealt{ven12,gei13}). The sdB in CD--30$^{\circ}$11223 will overflow its Roche Lobe in $\approx40$\,million years when the system will have shrunk to an orbital period of \porb$\approx 40$\,min. The sdB will start accretion onto the WD companion and after accreting $\approx 0.1$\,\msol, helium burning is predicted to be ignited unstably in the accreted helium layer on the surface of the WD \citep{bro15,bau17}. This could either disrupt the WD even if the mass is significantly below the Chandrasekhar mass, a so-called double detonation type Ia supernova (e.g. \citealt{liv90,liv95,fin10,woo11,wan12,she14,wan18}), or just detonate the He-shell without disrupting the WD; the latter scenario results in a faint and fast .Ia supernova with subsequent weaker He-flashes \citep{bil07,bro15}. Therefore, systems like CD--30$^{\circ}$11223 are predicted to be either the progenitors for double detonation type Ia supernovae or perhaps faint and fast .Ia supernovae when the WD is not disrupted. If the WD explodes, the companion will move translationally with the original orbital velocity, and US\,708 is a candidate which is proposed to be the former donor star in such a binary system \citep{gei15}. 

Sub-Chandrasekhar mass SN\,Ia explosions with thick helium shells accreted from a helium rich companion are predicted to exhibit an early time flux excess and red colors \citep{pol19}.  \citet{de19} discovered a Ia supernova which matches all the properties predicted for a double detonation SN\,Ia with a massive helium shell and pre-explosion properties similar to CD$-30^\circ11223$. %However, \citet{tow19} showed that double detonation type Ia supernovae with thin, modestly enriched helium layers can produce normal type Ia supernovae.

Although a clean picture has evolved with evidence for systems that will start accretion before the hot subdwarf has evolved into a WD, no accreting WD with a Roche lobe-filling hot subdwarf donor has yet been found. The only known hot subdwarf binary showing signs of accretion is HD\,49798, which consists of a massive sdO primary with a massive compact companion in a $1.55$ day orbit \citep{tha70, duf72, kud78}. \citet{isr95, isr97} reported X-ray pulses with a period of 13.2\,sec which was interpreted as the spin period of a magnetic compact companion accreting from the sdO wind. \cite{mer09} detected an eclipse in the X-ray light curve with a period coincident with the spectroscopic period, allowing them to derive precise masses of the sdO ($M_{\rm sdO}=1.50\pm0.05$\,\msol) and its compact companion ($M_{\rm }=1.28\pm0.05$\,\msol). Follow-up studies could not fully resolve the nature of the compact companion \citep{mer09,mer11,mer13,mer16}. \citet{bro17} showed that the massive sdO in HD\,49798 will fill its Roche Lobe in $\approx65$\,kyrs and start accretion onto its compact companion with a rate of $\approx10^{-5}$M$_\odot$yr$^{-1}$. 

Here, we present the discovery of the most compact sdOB binary ZTF\,J213056.71+442046.5 (hereafter \ztf) that matches the properties of a hot subdwarf star which fills its Roche lobe and has started mass transfer to the WD companion. We discovered the object in fall 2018, but did not announce our finding, in a search for periodic objects in the hot subdwarf catalog of $\approx40,000$ hot subdwarf candidates presented in \citet{gei19} using data from our dedicated high-cadence survey at low Galactic latitudes as part of the Zwicky Transient Facility. Independently, on 2019-05-25, Gabriel Murawski reported the object to the International Variable Star Index (VSX) \footnote{https://www.aavso.org/vsx/index.php?view=detail.top\&oid=689728} and in two Astronomers Telegrams, \citet{riv19} report the absence of X-ray emission from 1\,ks {\it Swift} observations and \citet{ram19} report spectroscopic observations suggesting a He-sdOB star classification for \ztf.  % The object was detected in a dedicated Galactic Plane high-cadence survey conducted with the Zwicky Transient Facility (Kupfer et al. in prep). 

\begin{table*}[t]
 \centering
 \caption{Summary of the observations}
  \begin{tabular}{lcllll}
  \hline\hline
 Date &    UT  &  Tele./Inst. & N$_{\rm exp}$ & Exp. time (s) & Coverage (\AA)/Filter \\
  \hline
  \multicolumn{2}{l}{{\bf Photometry}}   &   &  &  &     \\
2018-04-10 - 2018-11-21 &    & Palomar 48-inch  &  759  & 30 &  ZTF-$r$ \\
2018-03-31 - 2018-11-09  &    & Palomar 48-inch  &  175  & 30 &  ZTF-$g$ \\
2019-04-28  &   10:56 - 11:45   & 84-inch/KPED & 1452 & 2 & $g^\prime$ \\     
2019-05-01  &   10:33 - 11:49   & 84-inch/KPED & 2270 & 2 & $g^\prime$ \\     
2019-05-14  &   09:29 - 10:29   & 84-inch/KPED & 2950 & 1 & $g^\prime$ \\     
2019-05-28  &   09:30 - 11:30   & 84-inch/KPED & 3585 & 2 & $g^\prime$ \\    
2019-07-08  &  04:04 - 04:51  & GTC/HiPERCAM & 789/1577 & 3.54/1.77 & $u_s / g_s,r_s,i_s,z_s$ \\
     \noalign{\smallskip}
 \multicolumn{2}{l}{{\bf Spectroscopy}}   &   &  &  &     \\
2019-01-27  &  04:54 -  05:22 &  Keck/LRIS & 14 & 60    &  3200 - 5300 \\  %(R300B/R316R)   \vspace{-0.1cm}
2019-05-30  &  09:09 -  09:44  & 200-inch/DBSP  & 8 & 240    & 3500 - 10\,500  \\%(R300B/R316R)
2019-06-25  &  01:30 -  05:37  & WHT/ISIS  & 105 & 120    & 3100 - 5300 \& 6350 - 8100  \\%(R300B/R316R)
2019-06-26  &  02:12 -  05:26  & WHT/ISIS  & 86 & 120    & 3100 - 5300 \& 6350 - 8100  \\%(R300B/R316R)
 \noalign{\smallskip}
%  2010-02-18  & VLT/X-Shooter  &  1 & 300  \\ 
   \hline
\end{tabular}
\label{observ}
\end{table*}

%  was the first system which shows all properties consistent with a double detonation progenitor. The system will start accretion in about 40 Myr years when the sdB is still doing He-core burning \citep{gei13}. Since then we discovered OW\,J0741 which is an sdO with a period of 44min \citep{kup17a}. However, it turned out that the sdO in this system is likely not a helium burning star but a low-mass helium white dwarf cooling down. The contraction of the object due to cooling is faster than the orbital decay. Therefore, this system is likely to get in contact only at a period of $\approx$5\,min.
 
%The question remains, where are the accreting progenitor systems. White dwarfs are expected to accrete from an sdB for millions to tens of millions of years until enough mass is accumulated and the white dwarf explodes. 
 
%In the following I present a few notes on a peculiar hot subdwarf which I discovered as part of my search for periodic objects in the hot subdwarf catalog of $\approx40\,000$ hot subdwarf candidates presented in \citet{gei19}. My personal most favorable interpretation is, that this object could be the first sdO in an accreting system.

\section{Observations}

As part of the Zwicky Transient Facility (ZTF), the Palomar 48-inch (P48) telescope images the sky every clear night \citep{bel19,gra19}. %\mwc{I understand differentiating ZTF between the telescope and camera, but it still reads weird... so maybe say ZTF observing strategy?} 
\ztf\, was first discovered in a dedicated high-cadence survey at low Galactic latitudes performed in the ZTF-$r$ band \citep{bel19a}. The high-cadence data was complemented by data from the ZTF public survey obtained in 2018 which was made available after data release 1 on May 8, 2018. The ZTF light curves of \ztf\, consist of 759 epochs in the ZTF-$r$ band and 175 epochs in the ZTF-$g$ band. Image processing and light curve extraction of ZTF data is described in detail by \citet{mas19}.  

High-cadence follow-up observations with 1 and 2 sec exposure time were conducted using the Kitt Peak 84-Inch Electron Multiplying Demonstrator (KPED; \citealt{cou19}), which is a photometer that uses a frame-transfer, EMCCD to achieve 15\,ms dead time covering a $4.4\times4.4$\,arcmin field of view. Data reduction was carried out with a customized pipeline. All frames were bias-subtracted and flat-fielded. 

Additionally, \ztf\, was observed with HiPERCAM, a five-beam imager equipped with frame-transfer CCDs allowing the simultaneous aquisition of $u_s$, $g_s$, $r_s$, $i_s$ and $z_s$\footnote{HiPERCAM uses high-throughput versions of the SDSS filters known as Super-SDSS filters, and we denote these filters $u_s$, $g_s$, $r_s$, etc.} band images at  a rate of up to 1000 frames per second \citep{dhi16, dhi18}. For these data, HiPERCAM was mounted on the 10.4\,m Gran Telescopio Canarias (GTC) on the island of La Palma in Spain.  \ztf\, was observed on the night of 2019-07-07 at 1.766\,s cadence with a dead time of 10\,ms for 1576 frames with HiPERCAM, in a run lasting  46~min, covering a little more than one 39\,min binary orbit; the cadence in $u_s$ alone was a factor of 2 slower than the other four bands to optimise signal-to-noise. The sky was clear with $0.8$--$1.5"$ seeing. The data were reduced using the dedicated HiPERCAM pipeline\footnote{https://github.com/HiPERCAM/}, including debiasing and flat-fielding. Differential photometry was performed. 

Phase-resolved spectroscopy of \ztf\, was obtained using the Keck\,I Telescope and the blue arm of the Low Resolution Imaging Spectrometer (LRIS; \citealt{mcc98}) using a low resolution mode ($R\approx1400$). We obtained a total of 14 spectra. Data reduction was performed with the Lpipe pipeline\footnote{http://www.astro.caltech.edu/\~dperley/programs/lpipe.html}\citep{per19}. 

Optical spectra were also obtained with the Palomar 200-inch telescope and the Double-Beam Spectrograph (DBSP; \citealt{oke82}) using a low resolution mode ($R\approx1200$). Both arms of the spectrograph were reduced using a custom \texttt{PyRAF}-based pipeline \footnote{https://github.com/ebellm/pyraf-dbsp}\citep{bel16}. The pipeline performs standard image processing and spectral reduction procedures, including bias subtraction, flat-field correction, wavelength calibration, optimal spectral extraction, and flux calibration. 

Additionally \ztf\, was observed with the 4.2m William Herschel Telescope (WHT) and the ISIS spectrograph \citep{car93} using a low/medium resolution mode (R300B and R600R grating, $R\approx1500$ and $R\approx3500$). Ten bias frames were obtained to construct an average bias frame and 10 individual lamp flat-fields were obtained to construct a normalized flat-field. One dimensional spectra were extracted using optimal extraction and were subsequently wavelength and flux calibrated. An arc lamp spectrum was taken at the position of the target before and after each observing sequence for LRIS, DBSP and ISIS as well as after every hour for ISIS to account for telescope flexure.

All times in each data set were converted to the barycentric dynamical timescale, corrected to the solar system barycenter, MJD(BTDB). Table\,\ref{observ} gives an overview of all observations and the instrumental set-ups.

\begin{figure*}
\begin{center}
\includegraphics[width=0.68\textwidth]{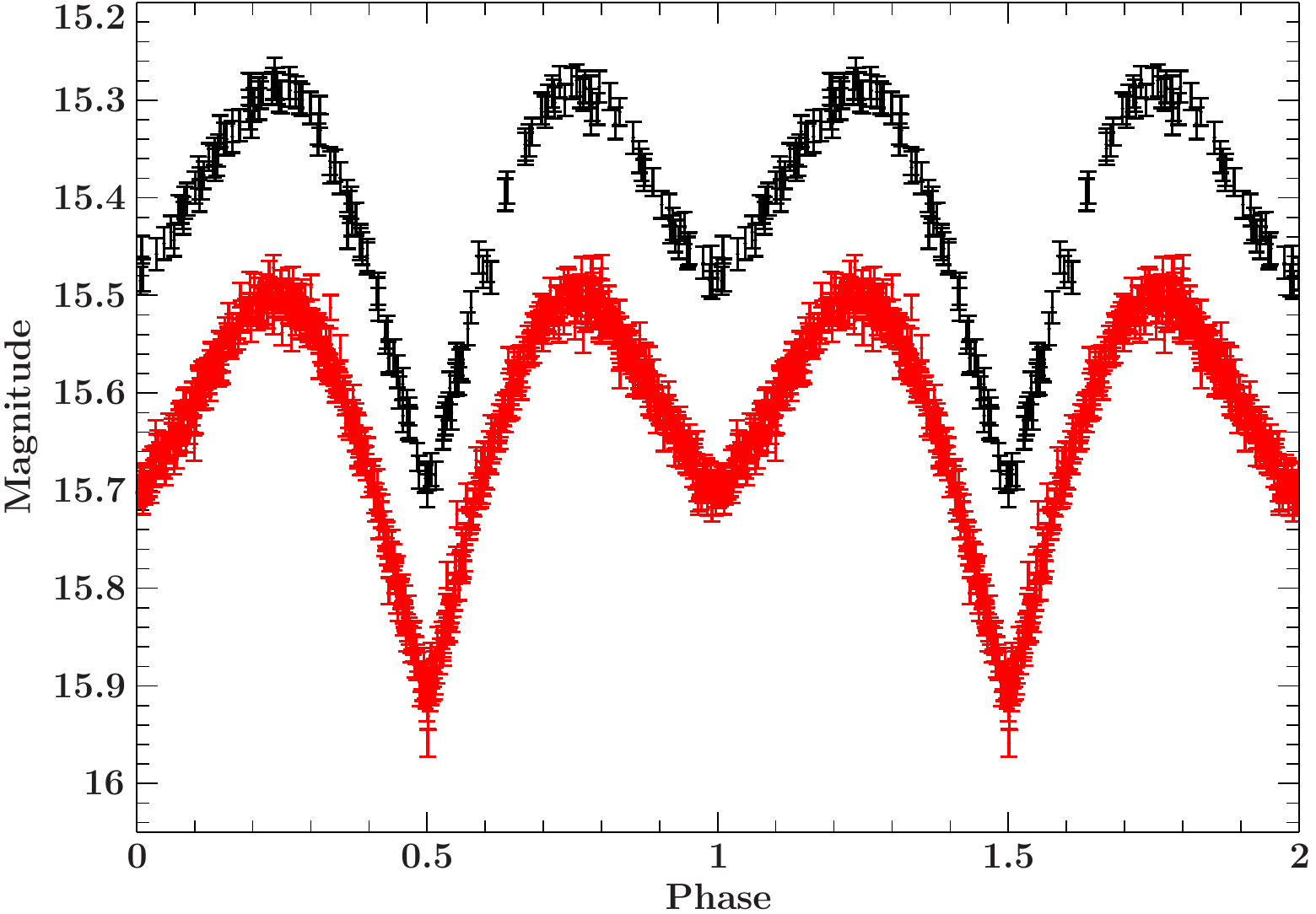}
\end{center}
\caption{Phase folded ZTF-$g$ (black) and ZTF-$r$ (red) discovery lightcurves.}
\label{fig:ztf_light}
\end{figure*}

\begin{figure*}
\begin{center}
\includegraphics[width=\textwidth]{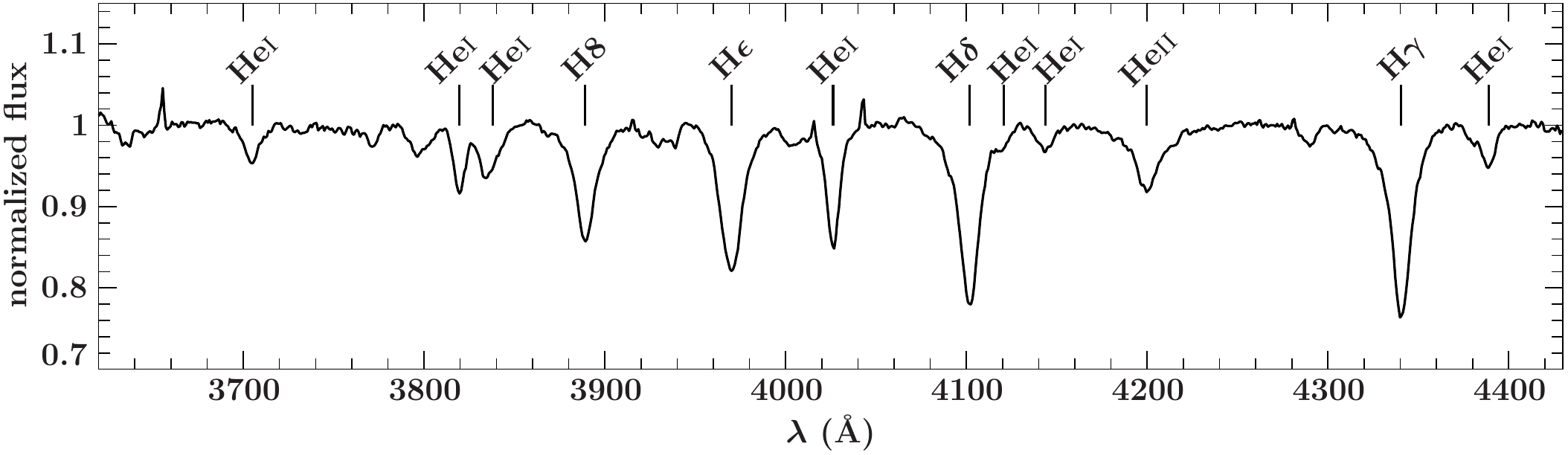}
\includegraphics[width=\textwidth]{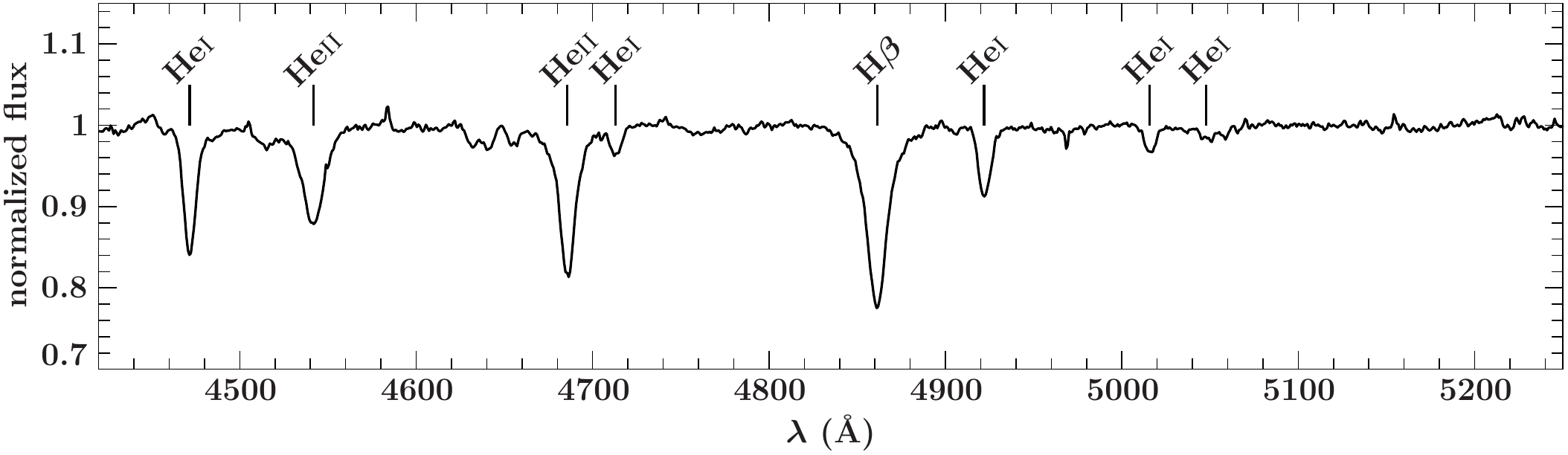}
\end{center}
\caption{Normalized average WHT spectrum of \ztf. All prominent lines are marked.}
\label{fig:ztf2130_aver}
\end{figure*}

\section{Orbital and atmospheric parameters}\label{orb_atm_pars}

\ztf\, shows strong periodic variability in its light curves (Fig.\,\ref{fig:ztf_light}). This variability is primarily caused by the tidal deformation of the sdOB primary under the influence of the gravitational force of the companion. We use the ZTF light curve with its multi-month baseline in combination with the KPED light curves and the HiPERCAM light curve to derive the orbital period of the system. The analysis was done with the \texttt{Gatspy}\footnote{http://dx.doi.org/10.5281/zenodo.14833}\citep{van15, van15a} module for time series analysis which uses the Lomb-Scargle periodogram \citep{sca82}. We find an orbital period of $P=39.3401(1)$\,min. The uncertainty was derived from a bootstrapping analysis. To determine the ephemeris, we measured the deepest point in the light curve, corresponding to the phase when the sdOB is furthest away from the observer. We used the parameter from the best model derived in section\,\ref{sec:lightcurve} and only fit the zero point of the ephemeris ($T_0$) to the HiPERCAM and KPED light curves. We find an ephemeris of:

\begin{equation}\label{equ:ephem}
%\begin{aligned}
T_{o} ({\rm BMJD}) = 58672.18085(78) \\
+ 0.0273195(2)E \noindent  
%\end{aligned}
\end{equation}

\begin{figure}
\begin{center}
\includegraphics[width=0.98\textwidth]{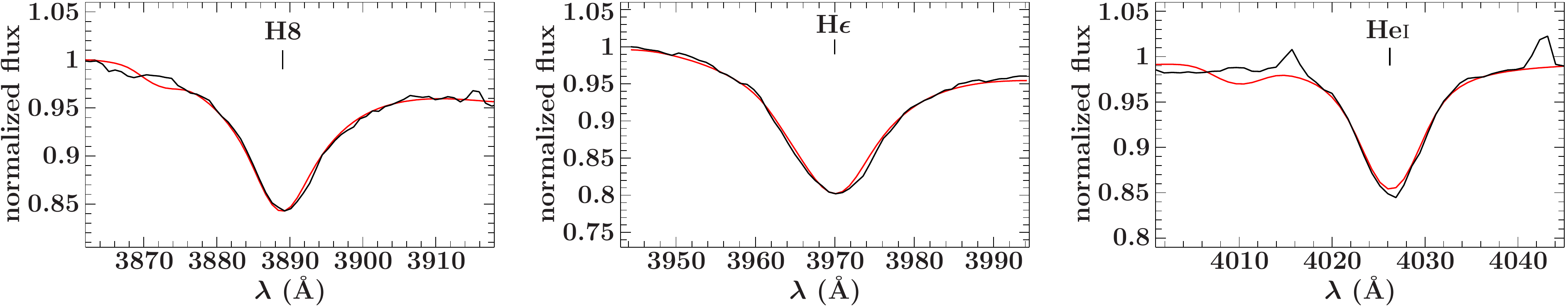}
\includegraphics[width=0.98\textwidth]{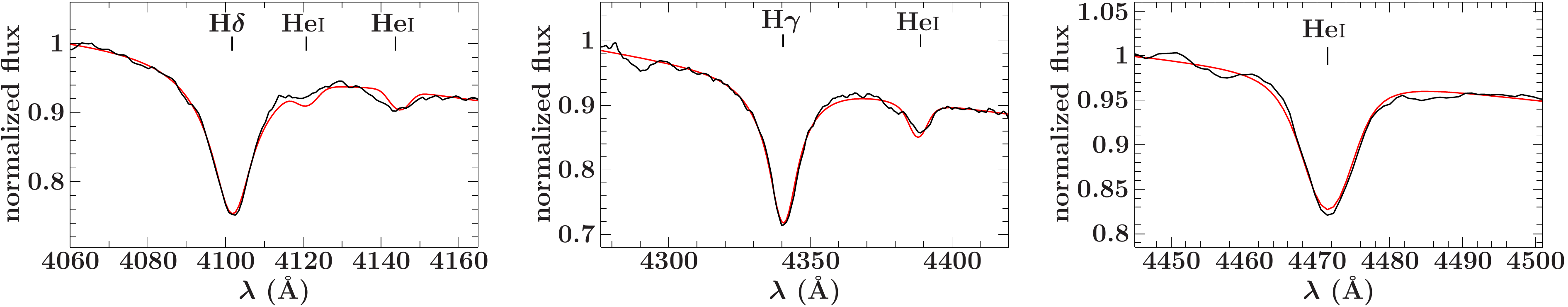}
\includegraphics[width=0.98\textwidth]{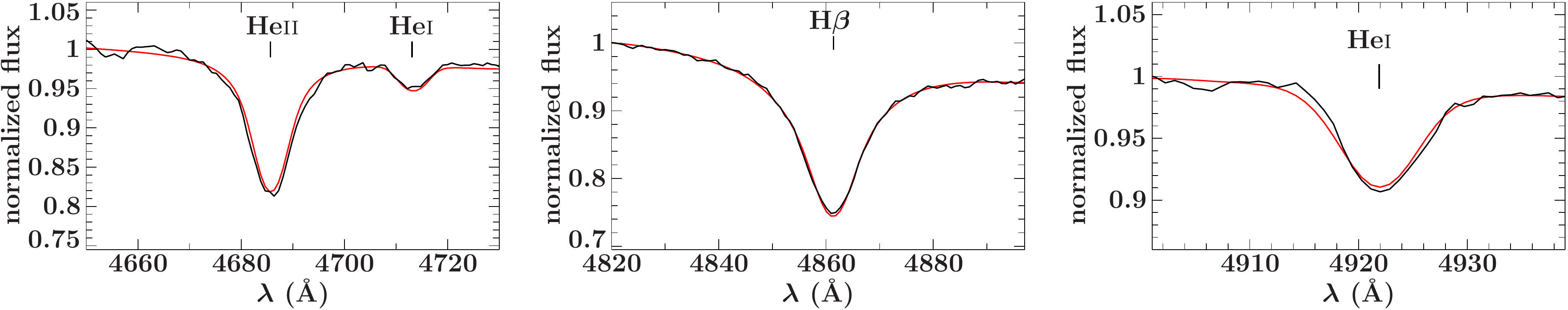}
\end{center}
\caption{Fit of synthetic NLTE models to the hydrogen Balmer, as well as neutral and ionized helium lines of the co-added WHT spectrum. The black solid line corresponds to the spectrum and red solid line to the fit.}
\label{fig:ztf2130_spec}
\end{figure} 

where $E$ corresponds to the epoch. 

The overall spectrum of \ztf\, shows Balmer lines as well as neutral (He\,{\sc i}) and ionized (He\,{\sc ii}) helium lines (Fig.\,\ref{fig:ztf2130_aver}); it can be well fit with a single He-sdOB star. We do not find evidence for a companion or accretion disk in the overall spectrum. The atmospheric parameters of effective temperature \teff, surface gravity, \logg\, and helium abundance, $\log{y}$, where $y=n({\rm He})/n({\rm H})$, and projected rotational velocity, \vrot, were determined for the sdOB by fitting the rest-wavelength corrected average LRIS, DBSP and WHT spectra with metal-free NLTE model spectra \citep{str07}. The ionization equilibrium between the He\,{\sc i} and He\,{\sc ii} is most sensitive to the effective temperature of the sdOB, whereas the broad hydrogen lines in the blue are most sensitive to $\log{g}$. The full procedure is described in detail in \citet{kup17,kup17a}. The best fit was derived using a $\chi^2$-minimization for each spectrum. We adopted the weighted mean of \teff, \logg, $\log{y}$ and \vrot\, as the final solution. We find \teff=$42,400\pm300$\,K, \logg=5.77$\pm$0.05 and $\log{y}$=$-$0.52$\pm$0.03. Figure \,\ref{fig:ztf2130_spec} shows the fit to the WHT spectrum and Tab.\,\ref{tab:atmo1} summarizes the results from the spectroscopic fits to the average LRIS, DBSP and WHT spectra. The occurrence of both He\,{\sc i} and He\,{\sc ii} as well as the increased helium abundance $\log{y}>-1$ classifies the hot subdwarf as an intermediate He-sdOB (see \citealt{heb16}). 

\begin{figure}
\begin{center}
\includegraphics[width=0.68\textwidth]{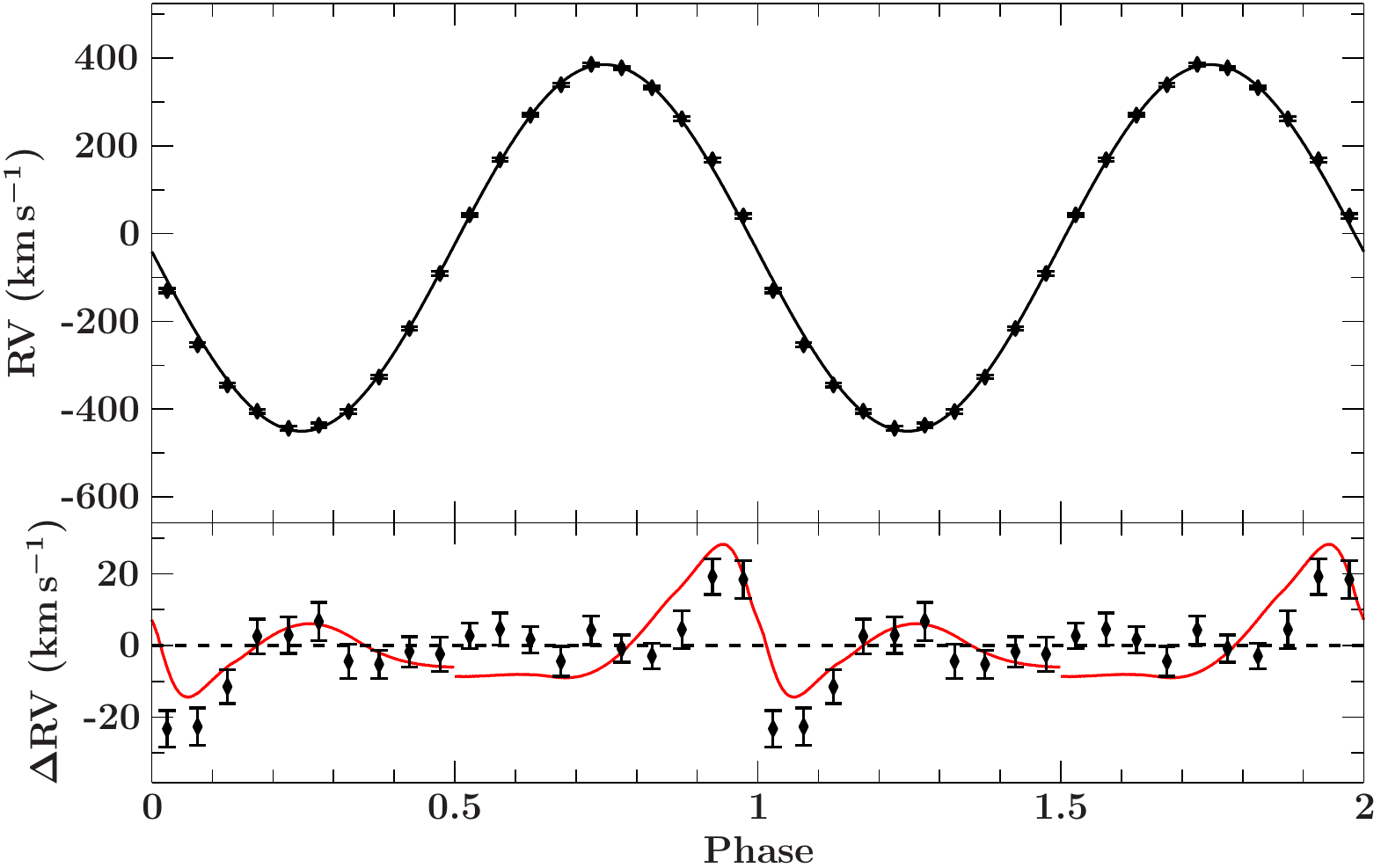}
\end{center}
\caption{Measured radial velocity measured versus orbital phase for \ztf. The RV data was phase folded with the orbital period and are plotted twice for better visualization. The residuals are plotted below. The strong deviation from a pure-sine curve around phase 0 (1) can be explained by the Rossiter-McLaughlin effect occurring when the accretion disk eclipses the rapidly rotating sdOB. The red curve shows the predicted residuals for the Rossiter-McLaughlin effect from our best fitting model (see section\,\ref{sec:lightcurve}). The RVs were measured from spectra obtained with WHT/ISIS.} %\trmc{I am wondering, given that the R-M effect modelling depends upon the light curve model, whether we shouldn't present this after the light curve data? I realise that we will need to forward refer to the RV amplitude, but the R-M effect stuff is complex to take in at this point.}
\label{fig:rv_curve1}
\end{figure}

To measure radial velocities (RVs), we folded the individual WHT/ISIS spectra on the ephemeris shown in equation (\ref{equ:ephem}) into 20 phase-bins and co-added individual spectra observed at the same binary phase. This leads to a signal-to-noise ($SNR$) per phase-bin of $\approx$100, well suited to measure velocities with a precision of $\approx$5\,\kms. Radial velocities were measured by fitting Gaussians, Lorentzians and polynomials to the hydrogen and helium lines to cover continuum, line and line core of the individual lines using the \texttt{FITSB2} routine \citep{nap04a}. The procedure is described in full detail in \citet{gei11a}. We fit the wavelength shifts compared to the rest wavelengths using a $\chi^2$-minimization. Assuming circular orbits, a sine curve was fit to the folded radial velocity (RV) data points (Fig.\,\ref{fig:rv_curve1}) excluding the data points around phase 0.8 - 1 and 0 - 0.2. We find a velocity semi-amplitude $K=418.5\pm2.5$\,\kms. Around phase 0 (or 1), when the sdOB is furthest away from the observer, the velocity curve deviates significantly from a pure sine-curve which can be explained with the Rossiter-McLaughlin effect when the rapidly rotating sdOB is eclipsed by the accretion disk. The red curve in Fig.\,\ref{fig:rv_curve1} show the residuals predicted from the Rossiter-McLaughlin effect calculated from our best fitting model (see Sec.\,\ref{sec:lightcurve}). To do so we assumed that the equivalent widths of the photospheric lines were constant at points over the sdOB, so that the absolute flux in the lines simply varies with the local surface brightness. Parts of the star blocked by the accretion disk were ignored when computing line profiles.

\begin{table}[t]
 \centering
 \caption{Summary of atmospheric parameter \ztf}
  \begin{tabular}{lcccc}
  \hline\hline
  Instrument &    \teff\,(K)  &  \logg & $\log{y}$ & \vrot \\
                    & (K)     &      &     &   (\kms)  \\
  \hline
  LRIS      &    $42,400\pm500$ &  5.78$\pm$0.08   &  $-$0.48$\pm$0.06   & 246$\pm$28  \\
  DBSP  &     $42,100\pm600$  &  5.82$\pm$0.07   &   $-$0.56$\pm$0.06   & 238$\pm$25 \\ 
  ISIS  &     $42,500\pm400$  &  5.72$\pm$0.06   &   $-$0.52$\pm$0.04   & 234$\pm$21 \\ \hline
   adopted & $42,400\pm300$  & 5.77$\pm$0.05  &  $-$0.52$\pm$0.03  &  238$\pm$15  \\
   \hline
\end{tabular}
\label{tab:atmo1}
%\begin{flushleft}
%$^a$ fixed to the values derived from ESI and HIRES\\
%$^b$ fixed to the values derived from DBSP and ISIS
%\end{flushleft}
\end{table}

\section{Light curve modelling}\label{sec:lightcurve}

The \texttt{LCURVE} code was used to perform the light curve analysis \citep{cop10}. \texttt{LCURVE} uses grids of points to model the two stars. The shape of the stars in the binary is set by a Roche potential. We assume that the orbit is circular and that the rotation periods of the stars are synchronized to the orbital period. The flux that each point on the grid emits is calculated by assuming a blackbody of a certain temperature at the bandpass wavelength, corrected for limb darkening, gravity darkening, Doppler beaming and the reflection effect. In what follows, orbital phase zero is the conjunction phase when the sdOB is furthest away from Earth and is being eclipsed by the accretion disk. Additionally, we fix the orbital period to the value determined in section \ref{orb_atm_pars},  as well as the effective temperature (\teff), radial velocity amplitude ($K$), and surface gravity ($g$) of the sdOB star (see section \ref{orb_atm_pars}).

\begin{figure}
\begin{center}
\includegraphics[width=0.98\textwidth]{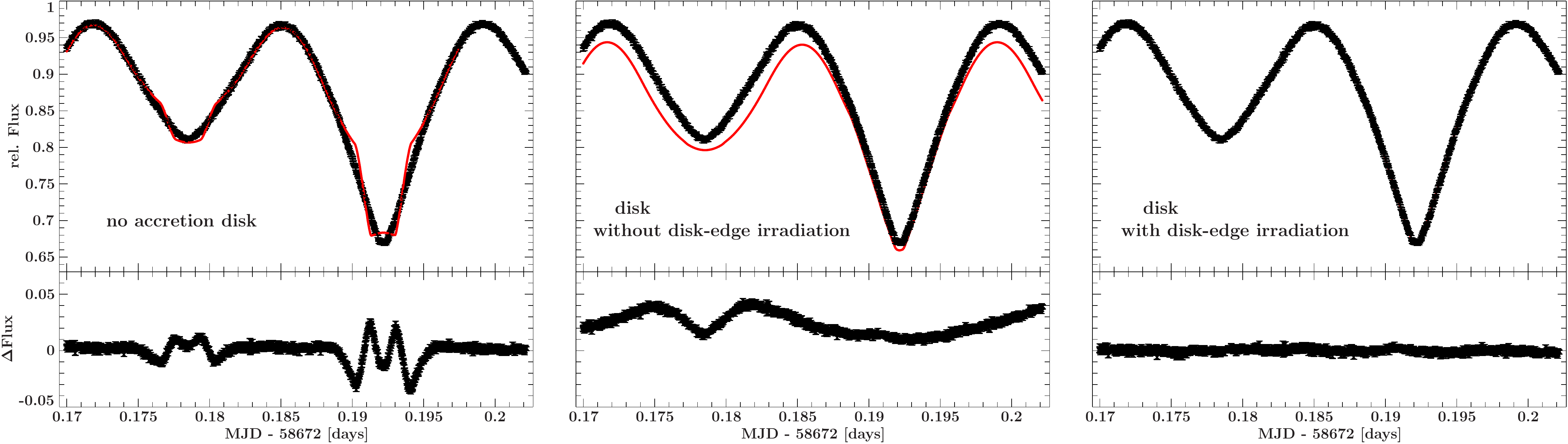}
\end{center}
\caption{The left panel shows the best fit (red curve) to the $g_s$-band HiPERCAM data (black points) using just two stars. The middle panel shows the fit without the flux from the irradiated rim of
the disk facing the sdOB star. The right panel shows the same fit when a disk and an irradiated rim of
the disk facing the sdOB star is added. The residuals are shown below}% \trmc{Are we going to show any of the other bands?}
\label{fig:modelling}
\end{figure}

The light curve of \ztf\ is dominated by ellipsoidal modulations due to tidal distortion of the sdOB. Ellipsoidal modulations are sensitive to the mass ratio, the size of the distorted star relative to the orbital separation and the limb and gravity darkening \citep{1985ApJ...295..143M}. We set the latter parameters according to the tables of \cite{cla11} shown in Tab.\,\ref{tab:fitparam}. We initially tried to model the light curve assuming that the system was composed of two detached stars, although we allowed the sdOB star to fill its Roche lobe if need be. It was quickly apparent that this model was inadequate as it shows large residuals around orbital phases 0 and 0.5 (left panel, Fig.~\ref{fig:modelling}). A value of $\chi^2 \approx 40,000$ was obtained for just $1576$ data points, an extremely poor fit.

The two-star model particularly fails around the phase when the sdOB is furthest from us. The light curve at this point shows a sharp and deep minimum that two stars alone cannot match. This is despite the sdOB star in our models expanding to fill $>99$\% of its Roche lobe, and thus maximizing the ellipsoidal modulations produced. The models were obtained through MCMC iteration, a highly robust form of optimisation since it is easy to ignore non-physical models such as overfilling of the Roche lobe. The only way the two-star model can even approach the data is to add a transit of the sdOB's companion star across the face of the sdOB to deepen the minimum at phase~0, which leads to a strong constraint upon the size of the companion star star as it must block $\approx 10$\% of the light from the sdOB. At phase~0.5, the eclipse of the companion star along with the radius constraint from phase~0, fixes the temperature of the companion. The resulting radius and temperature for the companion can be ruled out on astrophysical grounds, compounding the poor fit delivered by the two star model. The mass function from the radial velocity curve of the sdOB implies an absolute minimum mass for the companion of $0.22\,$\msol\, (for a zero-mass sdOB) and a more realistic minimum of $0.5\,$\msol\, if one assumes that $M_{\rm sdOB} \ge 0.25\,$\msol. The companion must therefore be a compact remnant, almost certainly a WD, as no other stars of this mass can fit within their Roche lobes at an orbital period of 39~min. However, the radius found for the companion is roughly four times larger than a WD of this mass, and its temperature is an implausibly low $\approx 2000\,$K. In summary, a simple two-star model cannot fit \ztf. We found the problem of the over-sized radius to be highly robust. Even allowing the limb and gravity darkening coefficients to float free (and to iterate towards implausible values), the radius of the sdOB's companion remained far too large for a WD of its mass.

The deep minimum at phase 0 points towards obscuration by some other structure. The compact nature of the binary and the tendency for the sdOB to fill its Roche lobe in the two-star model clearly suggests that we may be seeing occultation by an accretion disk associated with mass transfer. We therefore considered a model with two stars and an accretion disk. \texttt{LCURVE} allows for axi-symmetric disks with a height and temperature that varies with radius, and inner and outer radii. This immediately led to a much better fit, but one that was still some way off from the data with $\chi^2 \approx 15,000$ for just $1576$ data points. The main discrepancy was that the orbital phases of maximum flux in the data were closer to phase 0.5 when the sdOB is closest to us than the model seemed to be able to achieve. We then realised that if there is a disk in \ztf, it must be a highly unusual one in that it is strongly irradiated by the \emph{donor} star. This is, for example, in contrast to cataclysmic variable stars which have low luminosity, low-mass main-sequence donor stars and also compact object X-ray binaries which are strongly irradiated, but from sources located at the center of their disks. The configuration of \ztf\, means that the surface of the disk will be irradiated, quite possibly by more flux than is generated through accretion, and, moreover, the rim of the disk that is closest to the sdOB star will be particularly strongly irradiated. We therefore added a disk edge component in the form of a squat cylinder of radius and height equal to that at the outer radius of the accretion disk. We allowed this edge to have a fixed temperature supplemented by irradiation determined by flux balancing according to the level of irradiating flux each element receives from the sdOB star, modelled as a point source. The edge component is particularly hot (around $30,000\,$K) on the side of the disk closest to the sdOB, and hottest at the point where the disk edge crosses the line of centers between the two stars. This leads to a sinsuoidal modulation of flux which peaks at phase~0.5 (ignoring the possibility of eclipse for the moment), bringing the phases of maximum flux closer to 0.5 as observed. 

\begin{table}
 \centering
 \caption{Overview of the fixed parameters for the LCURVE fit to the HiPERCAM light curve for \ztf}
  \begin{tabular}{lrrrrr}
  \hline\hline
  Parameter &   $u_s$ & $g_s$ & $r_s$ & $i_s$ & $z_s$ \\
  \hline
  Beaming Factor (F)      &    1.55   &  1.40   &  1.30   & 1.24 & 1.20  \\
  gravity darkening $\beta$   &  0.34  &  0.28  &  0.27  & 0.26   & 0.26 \\
  limb darkening $a_\mathrm{1}$    &    1.20    &  1.22   &  1.17    &   1.10     &    1.12   \\
  limb darkening $a_\mathrm{2}$   &     --1.74   &  --1.82  & --1.89    &  --1.82      &   --1.82    \\
  limb darkening $a_\mathrm{3}$   &     1.56    &  1.62   &  1.70    &  1.64       &    1.72    \\
  limb darkening $a_\mathrm{4}$   &    --0.54    &  -0.55  &   --0.57   &   --0.56     &  --0.60      \\
     \hline
\end{tabular}
\label{tab:fitparam}
%\begin{flushleft}
%$^a$ fixed to the values derived from ESI and HIRES\\
%$^b$ fixed to the values derived from DBSP and ISIS
%\end{flushleft}
\end{table}

\begin{figure}
\hspace*{\fill}
\includegraphics[width=0.47\textwidth]{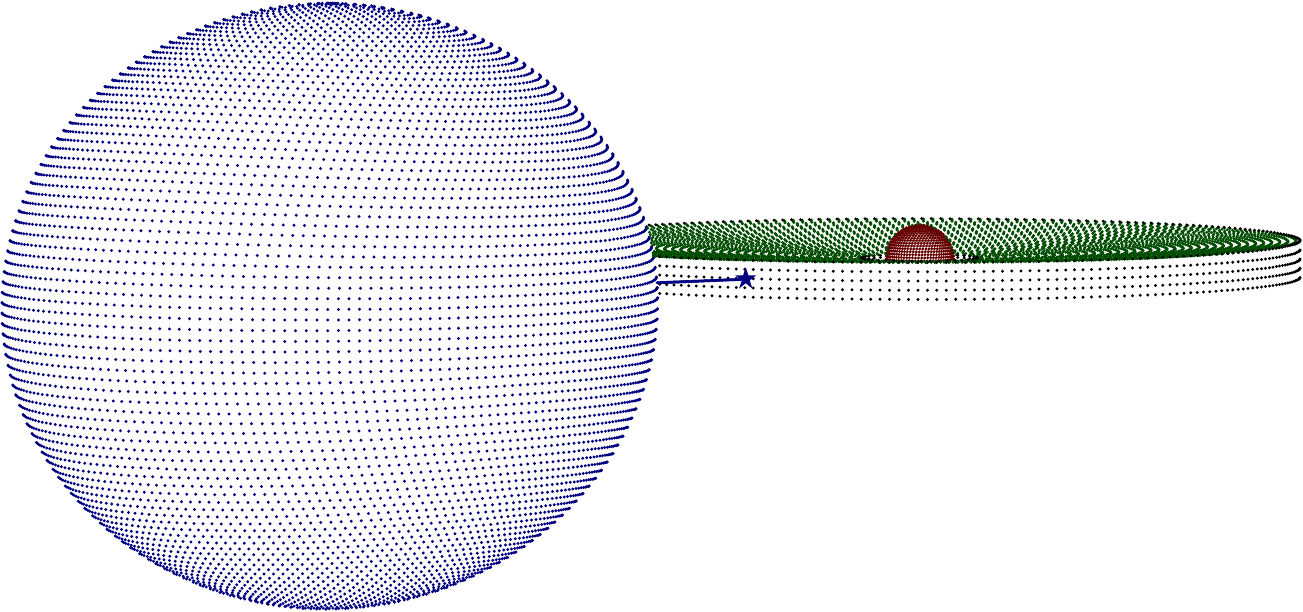}
\hspace*{\fill}
\includegraphics[width=0.47\textwidth]{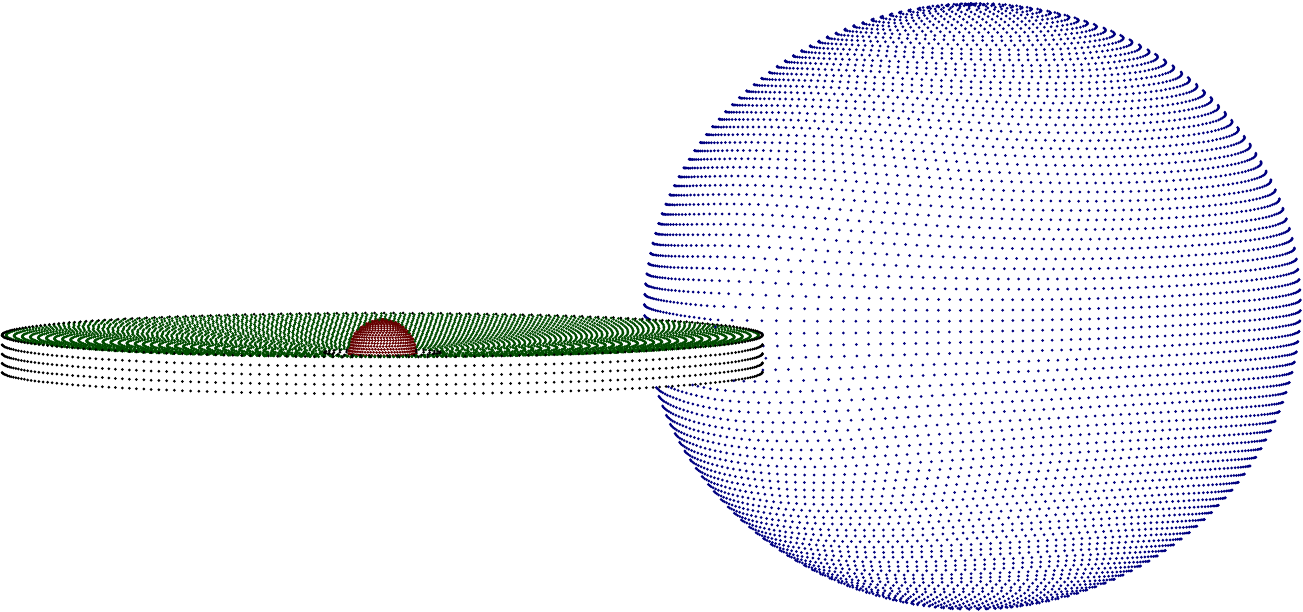}
\hspace*{\fill}
\caption{A visualisation of the grids used to model \ztf\ seen at
orbital phases $0.4$ (left) and $0.9$ (right). The actual grids used had a higher resolution than those displayed here.}
\label{fig:grids}
\end{figure}

\begin{table}
\centering
\caption{Overview of the measured and derived parameters for \ztf.}
\begin{tabular}{lll}
\hline\hline
Right ascension$^a$ & RA [hrs]  & 21:30:56.71 \\
Declination$^a$  & Dec $[^\circ]$  & 44:20:46.450 \\
Magnitude$^b$ & $g$ [mag] & 15.33$\pm$0.01 \\
Parallax$^a$  & $\varpi$    [mas] & 0.8329$\pm$0.0305 \\ 
Distance      &  $d$ [kpc] &  1.20 $\pm$0.06  \\
Absolute Magnitude   &  \multirow{2}{*}{$M_{\rm g}$ [mag]}   & \multirow{2}{*}{4.3$\pm$0.2}  \\
(reddening corrected) &   &  \\
Proper motion (RA) &   $\mu_\alpha$cos$(\delta)$  [mas\,yr$^{-1}$]   &  $0.009\pm0.047$ \\
Proper motion (Dec) &   $\mu_\delta$  [mas\,yr$^{-1}$]   &  $-1.682\pm0.048$  \\
\hline
\multicolumn{3}{l}{\bf{Atmospheric parameters of the sdOB}}  \\ 
Effective temperature & \teff\,[K] & $42,400\pm300$ \\
Surface gravity   & \logg  & $5.77\pm0.05$  \\
Helium abundance & $\log{y}$  & $-0.52\pm0.03$ \\
Projected rotational velocity & \vrot\,[\kms] &  238$\pm$15  \\
\hline
\multicolumn{3}{l}{\bf{Orbital parameters}}   \\ 

Ephemeris zero point &  $T_0$ [MBJD]  &  58672.18085(78) \\
Orbital period & \porb\,[min]  & 39.3401(1)  \\
RV semi-amplitude (sdOB) & $K$ [\kms] & 418.5$\pm$2.5 \\
System velocity & $\gamma$\,[\kms] & $-$33.9$\pm$1.9 \\ 
Binary mass function & $f_{\rm m}$ [\msol] & 0.2075$\pm$0.0037  \\
\hline
\multicolumn{3}{l}{\bf{Derived parameters}} \\

Mass ratio  &  $q = \frac{M_{\rm sdOB}}{M_{\rm WD}}$  & 0.617$\pm$0.015  \\
sdOB mass &  $M_{\rm sdOB}$ [\msol] & 0.337$\pm$0.015 \\ 
sdOB radius & $R_{\rm sdOB}$ [R$_{\odot}$] &  0.125$\pm$0.005 \\ 
WD mass &  $M_{\rm WD}$ [\msol] & 0.545$\pm$0.020 \\
Orbital inclination & $i\,[^\circ$] &  86.4$\pm$1.0  \\
Separation  & $a$ [R$_{\odot}$]   &  0.367$\pm$0.004 \\
\hline
\end{tabular}
\begin{flushleft}

$^a$ from Gaia DR2 \citep{gai18}\\
$^b$ from PanSTARRS DR1 \citep{cham16}
\label{tab:system}
\end{flushleft}
\end{table}

The irradiated disc-edge model is shown in the right panel of Fig.~\ref{fig:modelling}. The value of $\chi^2$ for this model was around 2000 for 1576 points, resulting in a reduced $\chi^2\approx1.3$, which is a considerable improvement over each of the other two models discussed. While there are still some residuals, they do not have the symmetry of the main light curve and presumably reflect variations in the geometry of the disk not captured in our model. We are more surprised by how small these residuals are than by their presence; there is for instance no obvious sign of a contribution from a ``bright-spot'' where the mass transfer stream hits the disc. Finally, in the middle panel of Fig.~\ref{fig:modelling}, we show the identical model but with the flux from disk edge turned off to show its significance and how it improves the agreement between the phases of maximum flux in the model compared to the data; in this case we do account for the eclipse of the disk edge by the sdOB.
%\mwc{I found this whole section really excellent}

\begin{figure*}
\begin{center}
\includegraphics[width=0.48\textwidth]{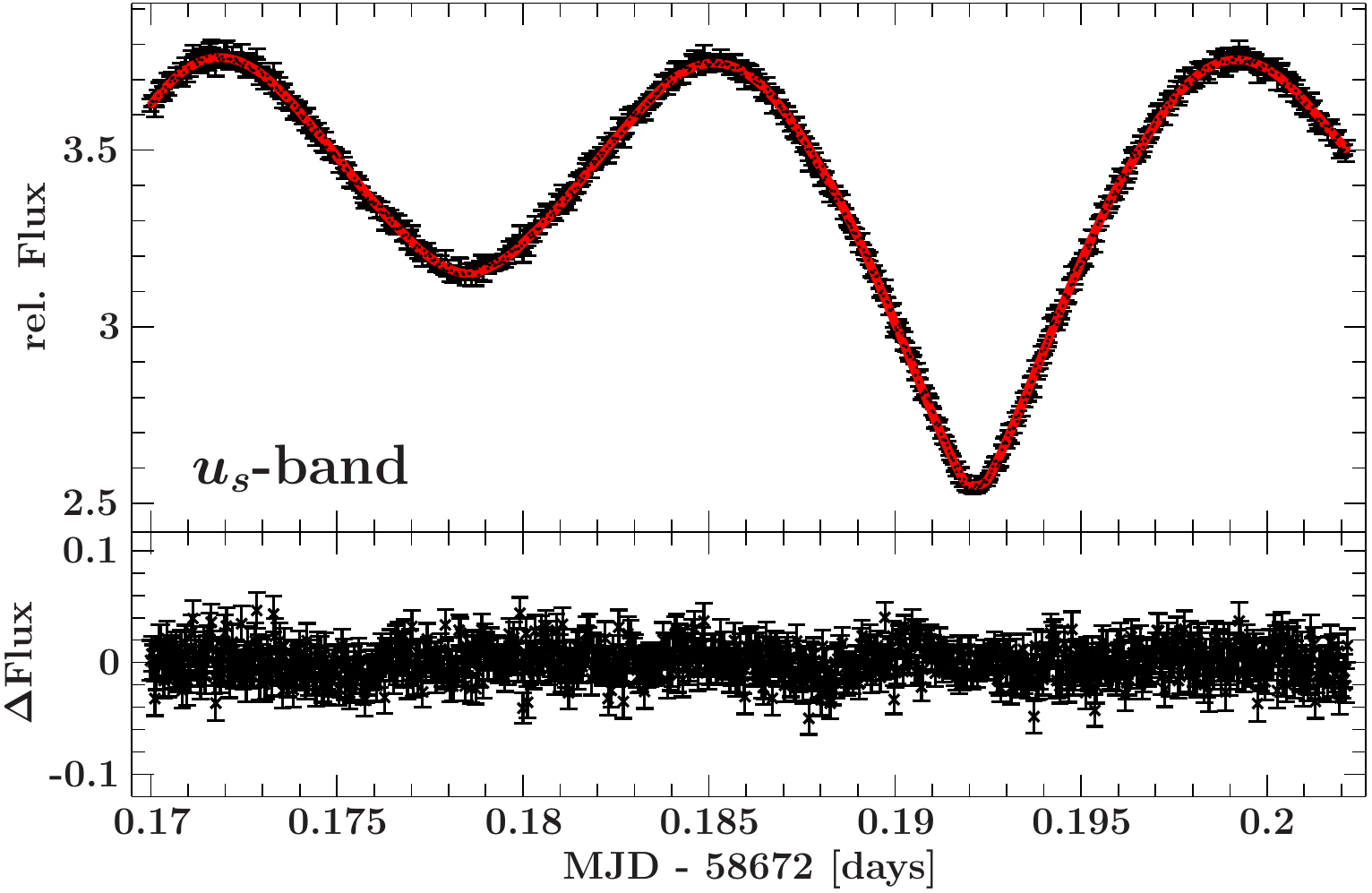}
\includegraphics[width=0.48\textwidth]{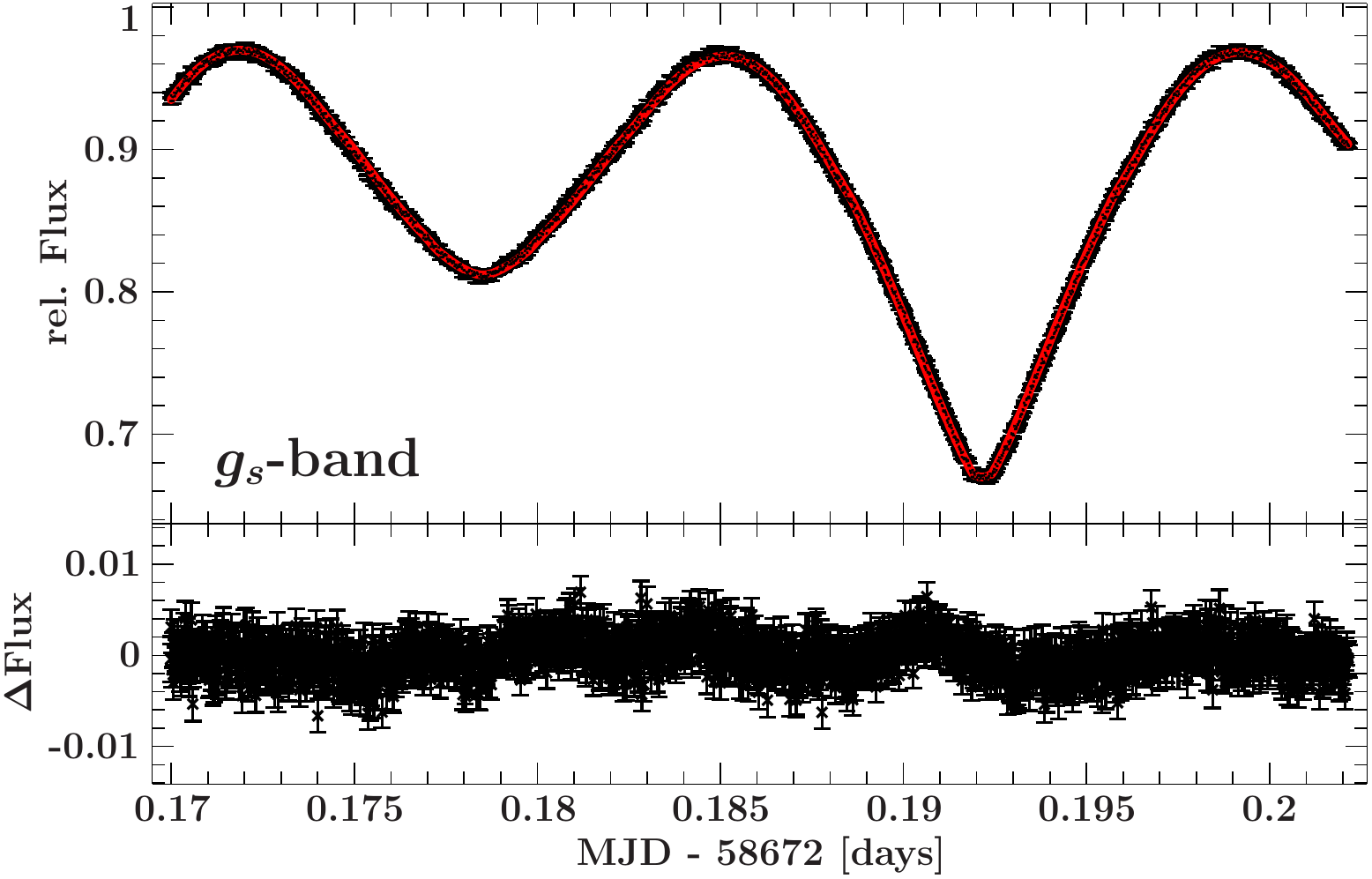}
\includegraphics[width=0.48\textwidth]{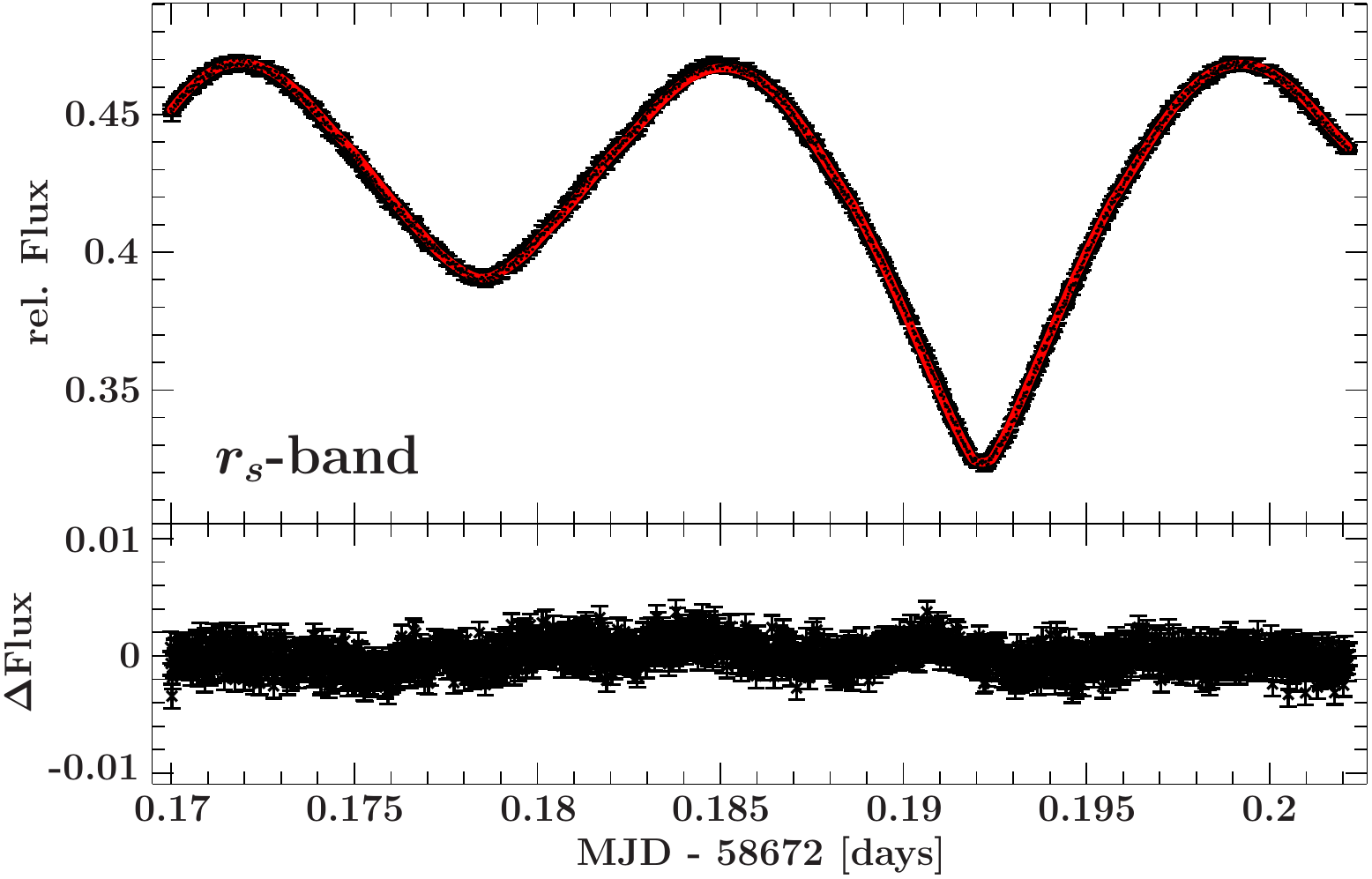}
\includegraphics[width=0.48\textwidth]{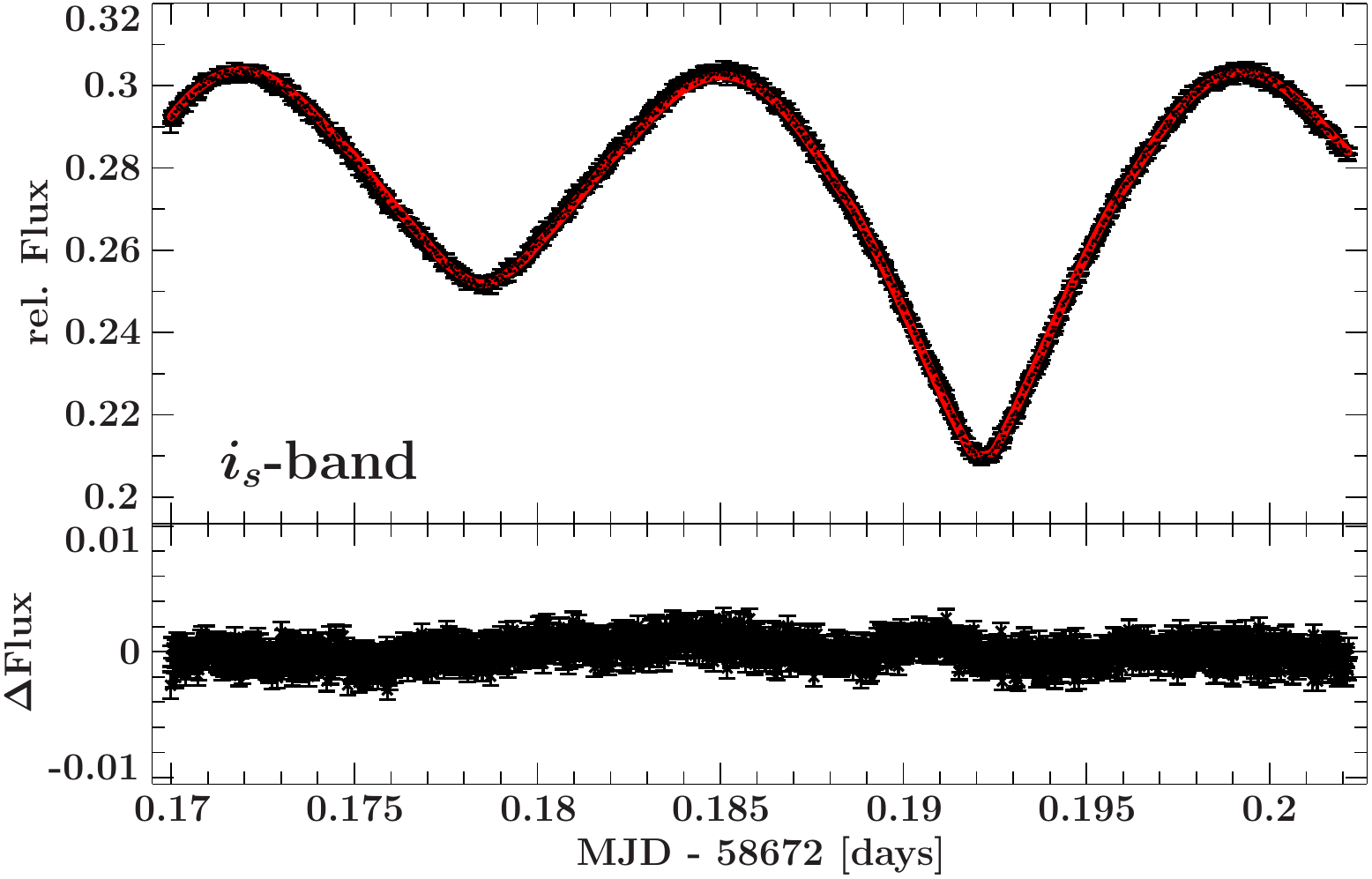}
\includegraphics[width=0.48\textwidth]{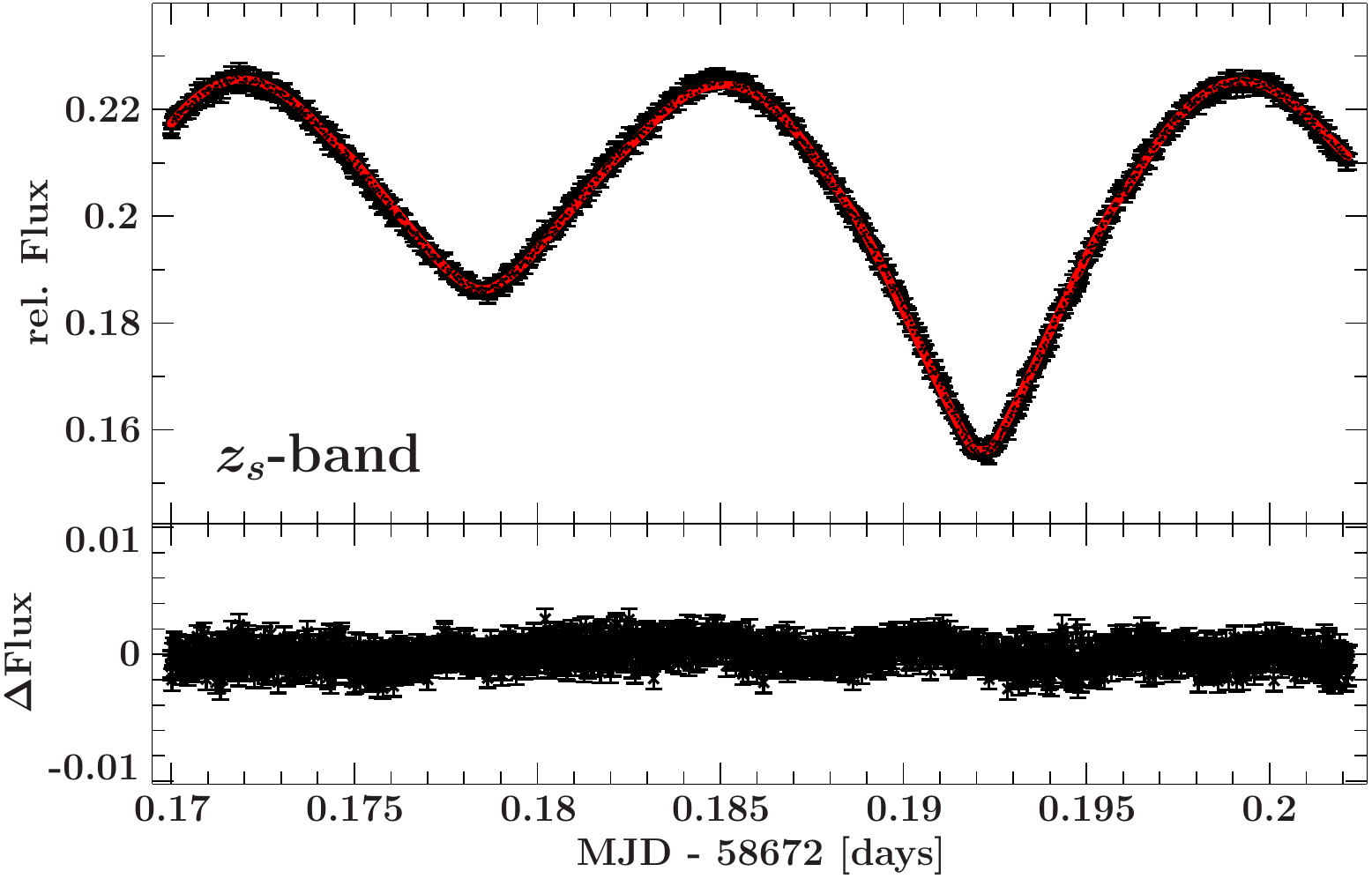}
\end{center}
\caption{HiPERCAM light curves (black) shown together with the \texttt{LCURVE} fits (red) for all five simultaneously observed optical Super-SDSS bandpasses. }
\label{fig:light_fits_2130}
\end{figure*}

For the final model we normalized the errors in the data to account for the small additional residuals and to obtain a reduced $\chi^2\approx1$. We assume a Roche lobe-filling sdOB star, an irradiated disk and an accreting WD. The passband specific beaming parameter $B$ ($F_\lambda = F_{0,\lambda} \lbrack 1 - B \frac{v_r}{c}\rbrack$, see \citealt{blo11}) was calculated following the approximation from \citet{loe03}. The passband specific gravity darkening and limb-darkening was taken from \citet{cla11} for a \teff$=42,500$\,K and \logg=5 star as higher \logg\, values are not available. We investigated how the gravity darkening ($\mathrm{\beta}$) and limb darkening coefficients ($a_\mathrm{1}$, $a_\mathrm{2}$, $a_\mathrm{3}$, $a_\mathrm{4}$) affect the results by adding them as free parameters with Gaussians around the theoretical value with FWHMs of $\sigma_\mathrm{\beta}=0.03$, $\sigma_{a_\mathrm{1}}=0.05$ and $\sigma_{a_\mathrm{1}}=0.05$. The variables $a_\mathrm{3}$, $a_\mathrm{4}$ were not varied. The co-variance between the gravity darkening and limb darkening parameter and system parameters is negligible compared to the uncertainty on the parameters. Therefore, we kept the limb darkening and gravity darkening coefficients fixed to the theoretical values from \citet{cla11}. The values used for the beaming, limb darkening and gravity darkening are shown in Tab.\,\ref{tab:fitparam}. We did not use any limb darkening or gravity darkening in the WD model, since these do not affect the light curve. This leaves as free parameters in the model the mass ratio $q$, the inclination $i$, WD temperature $T_{\rm WD}$, the scaled radius of the WD companion $r_{\rm WD}$, the velocity scale ($\mathrm[K+K_{\rm WD}]/\sin i$), the scaled disk size, disk height, disk temperature as well as the disk edge temperature and a bandpass-dependent disk edge reflection coefficient. Besides these system parameters we added a first order polynomial to correct for any residual airmass effects. 

\begin{figure*}
    \centering
    \includegraphics[width=0.85\textwidth]{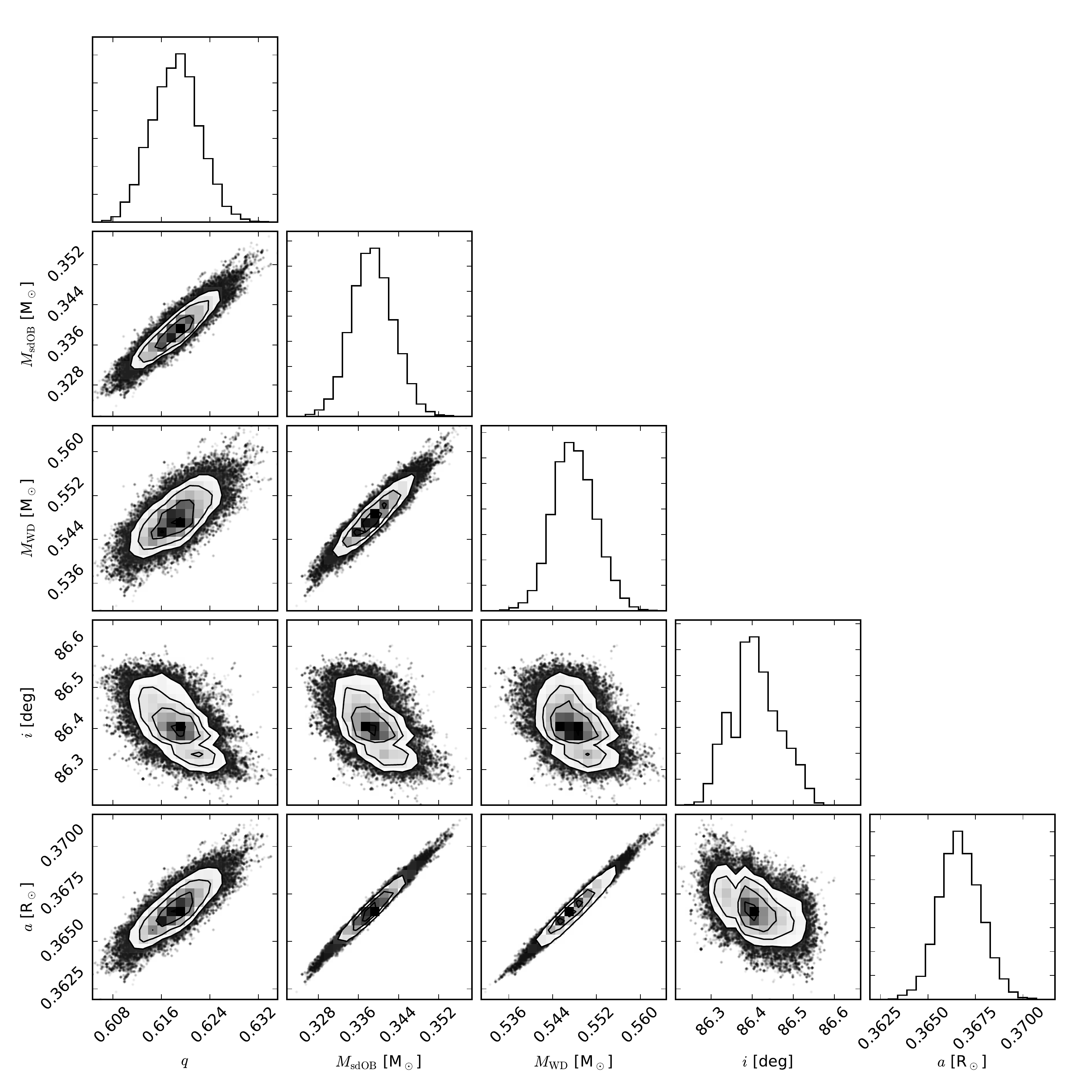} 
    \caption{Corner plots of the physical parameters inferred as a result of the analysis combining the light curve modelling with radial-velocity and spectral line fitting results.}
    \label{fig:corner}
\end{figure*}

To determine the uncertainties in the parameters, we combine \texttt{LCURVE} with \texttt{emcee} \citep{for13}, an implementation of an MCMC sampler that uses a number of parallel chains to explore the solution space. We used 256 chains and let them run until the chains stabilized to a solution, which took approximately 2000 generations. 

Figure~\ref{fig:grids} shows visualisations of the model grids, showing the approximate geometry and sizes of the various components adopted. The gas stream and bright-spot, although illustrated, were not used during model computations. Figure~\ref{fig:light_fits_2130} shows the fit for each HiPERCAM band and Fig.\,\ref{fig:corner} shows the corner plot for the best solution from the MCMC sampler.

%\emph{TRM: stuff to do: define all limb darkening, gravity darkening and Doppler beaming coefficients for the sdOB in all HiPERCAM bands.}

\section{System parameters}\label{sec:system}

The strong light curve variability caused by ellipsoidal modulation and the eclipsing accretion disk in combination with the radial-velocity amplitude and the spectral fits, allow us to derive system parameters. Solutions were calculated from a simultaneous fit to the five HiPERCAM light curves (Tab.\,\ref{tab:system}).

We find that the system consists of a low mass sdOB with a typical WD companion. The sdOB is Roche lobe-filling and has a volumetric-corrected radius of $R_{\rm sdOB}=0.124\pm0.005$\rsol. A mass ratio $q = M_{\rm sdOB}/M_{\rm WD}=0.617\pm0.015$, a mass for the sdOB  $M_{\rm sdOB}=0.337\pm0.015$\,\msol\, and a WD companion mass $M_{\rm WD}=0.545\pm 0.020$\,\msol\, were derived. The mass of the sdOB is significantly lower than for a canonical sdOB with $\approx0.48$\,\msol. The radius of the sdOB star is found to be typical for an sdOB star and the inclination is found to be $i=86.4\pm1.0$\,\degree\,(Tab.\,\ref{tab:system}). From the system parameters we find that sdOB would have a projected rotational velocity \vrot$=227\pm10$\,\kms\, if synchronized to the orbit. The measured \vrot$=238\pm15$\,\kms is consistent with a synchronized orbit.

We calculate the absolute magnitude ($M_{\rm g}$) of \ztf\, using the visual PanSTARRS g-band magnitude ($m_{\rm g}=15.33\pm0.01$\,mag; \citealt{cham16}) and the parallax from Gaia DR2 ($\varpi$=0.8329$\pm$0.0305\,mas; \citealt{gai16,gai18}). Because \ztf\, is located near the Galactic Plane, significant reddening can occur. \citet{gre19} present updated 3D extinction maps based on Gaia parallaxes and stellar photometry from Pan-STARRS 1 and 2MASS \footnote{http://argonaut.skymaps.info/}  and find towards the direction of \ztf\, an extinction of $E(B-V)=0.18\pm0.02$ at a distance of $1.2\pm0.06$\,kpc; this results in a total extinction in the g-band of $A_{\rm g}=0.63$\,mag. With the corrected magnitude, we find an absolute magnitude of $M_{\rm g}=4.3\pm0.2$\,mag, consistent with a hot subdwarf star \citep{gei19}.

%\trmc{Should point out that our model is fairly degenerate. e.g. I would not believe many of the details such as the precise inner and outer radii and inclination, because the disk model is crude. It could be partially transparent, its edge not vertical, not of uniform height, etc, etc. The things we can conclude I think are that something extra is needed to obscure the sdOB and that a simple disk model does a decent job. An early look probably shows that the masses for the two stars from the LC model come back rather small, so we should also consider a model in which we enforce a prior on the sdOB to raise its mass to the canonical value to see how much worse the model is. Can we get a prior on the radius from the Gaia parallax and spectrum fit? This could be very useful I think.

\section{Discussion}

%\subsection{Constraints from Gaia DR2}\label{sec:gaia}
%We estimate the absolute magnitude of the objects, using the distances from the Gaia DR2 parallaxes \citep{gai18}.

\subsection{Comparison with Gaia parallax}\label{sec:gaia}
To test whether our derived system parameter are consistent with the parallax provided by Gaia DR2 ($\varpi$=0.8329$\pm$0.0305\,mas), we compared the measured parameter from the light curve fit to the predictions using the Gaia parallax. The approach follows a similar strategy to that described by \citet{rat19}. Using the absolute magnitude $M_{\rm g}=4.3\pm0.2$\,mag, we calculate the luminosity:
\begin{equation}
L_{\mathrm{sdOB}}=L_{0}10^{-0.4(M_{\rm{g}}+BC_{\rm{g}})}.
\end{equation}
With $L_{0}= 3.0128\times10^{33}$\,erg\,s$^{-1}$ and a bolometric correction $BC_{\mathrm{g}}=-3.6$\,mag derived for our stellar parameters from the {\tt MESA} Isochrones \& Stellar Tracks (MIST; \citealt{dot16, cho16, pax11, pax13, pax15, pax18}), we find  $L_{\rm{sdOB}}=41\pm9$\,L$_\odot$. Using the Stefan-Boltzmann law applied to a black body ($L = 4 \sigma \pi R_{\rm{sdOB}}^{2}T_{\rm{eff}}^4$), we solve for the radius of the sdOB star, and combined with $R_{\rm{sdOB}}^{2}=GM_{\rm{sdOB}}/g$, we derive the the sdOB mass:
\begin{equation}
M_{\rm{sdOB}} = \frac{L_{\rm{sdOB}}10^{\log{g}}}{4\pi GT_{\rm{eff}}^4},
\end{equation}
giving $M_{\rm{sdOB}} = 0.30\pm0.08$\,\msol\, and a radius $R_{\rm{sdOB}} = 0.12\pm0.02$\,\rsol, both are in agreement with the results from the light curve and spectroscopic fits and confirm that this sdOB star is lower in mass than the canonical hot subdwarf stars. 

%. With the mass-radius relation (equation 3), we express the luminosity ($L = 4 \sigma \pi R^2T^4$ from the Stephan-Boltzman equation applied to a black body) as a function of mass and surface gravity instead of radius. Using the zero-point luminosity, we solve for the mass, combine constants, and simplify to the following formula:

\subsection{Evolutionary history}
\begin{figure}
    \centering
    \includegraphics[width=0.49\textwidth]{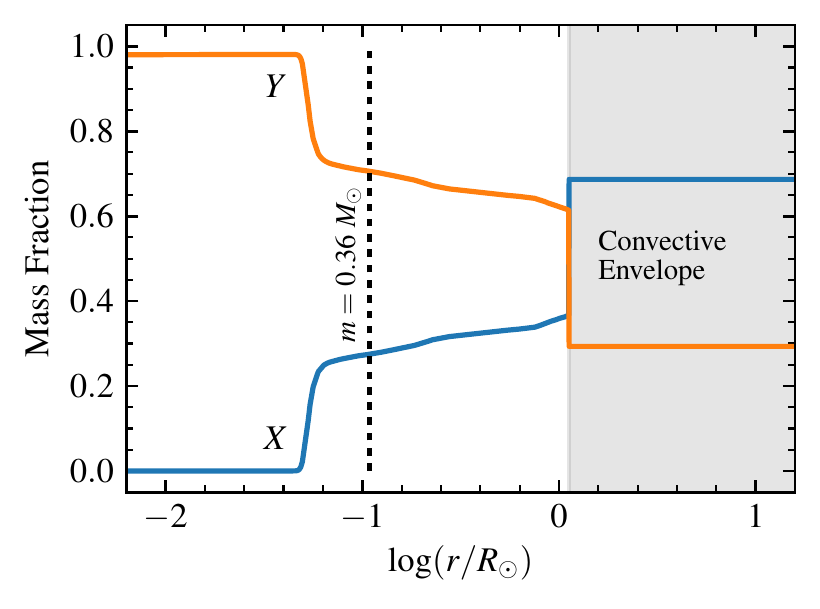}
    \caption{Hydrogen ($X$) and helium ($Y$) composition profile for a 2.8~$M_\odot$ RGB star just before core helium burning begins and outer material is removed to make a hot subdwarf. The black dashed line shows the location of mass coordinate $m=0.36\ M_\odot$ corresponding to the surface of the hot subdwarf after the envelope is removed. Nuclear burning has partially processed the material at this location due to a former convective core that receded over the duration of the main sequence.}
    \label{fig:internal-profile}
\end{figure}

In order to match the high $T_{\rm eff}$ observed with a hot subdwarf that still retains some surface hydrogen, we construct binary evolution models of hot subdwarfs that come into contact with a WD companion only after they have completed core helium burning and evolved toward hotter temperatures. We use {\tt MESA} version 12115 to construct these models \citep{pax11,pax13,pax15,pax18,pax19}. We start by constructing low-mass He-burning hot subdwarf models from progenitors with masses in the range $2.5\text{--}2.8\ M_\odot$, which have main sequence lifetimes of 400-500~Myr. After evolving onto the red giant branch (RGB), these stars achieve core temperatures high enough for non-degenerate helium ignition and hence can form He-burning cores with masses lower than the $0.48$\,\msol\, required for a degenerate He core flash. Figure~\ref{fig:internal-profile} shows the internal composition profile for a $2.8\ M_\odot$ RGB model as it begins central He burning. Main sequence stars in this mass range develop convective cores that initially encompass $\approx\,$0.5~$M_\odot$ but then recede to encompass less than 0.2~$M_\odot$ by the end of core hydrogen burning. After the He core forms and begins to grow as the star ascends the RGB, the region outside the He core reflects a composition that has been partially processed by nuclear burning, with H still present but less abundant than its primordial value. Because the convective core burning on the main sequence is dominated by the CNO cycle, we also expect that the interior region that eventually forms the hot subdwarf will be depleted of carbon and rich in nitrogen.

\begin{figure*}
    \centering
    \includegraphics[width=0.49\textwidth]{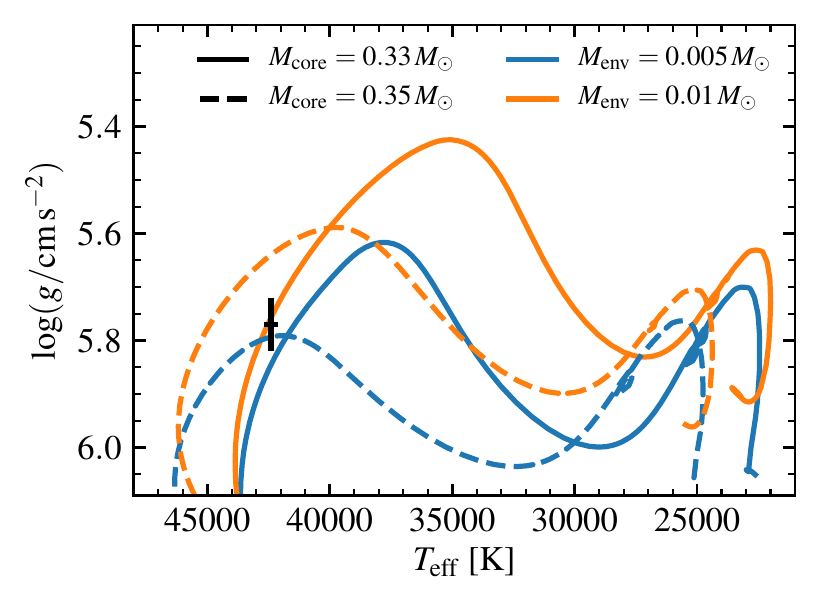} 
    \includegraphics[width=0.49\textwidth]{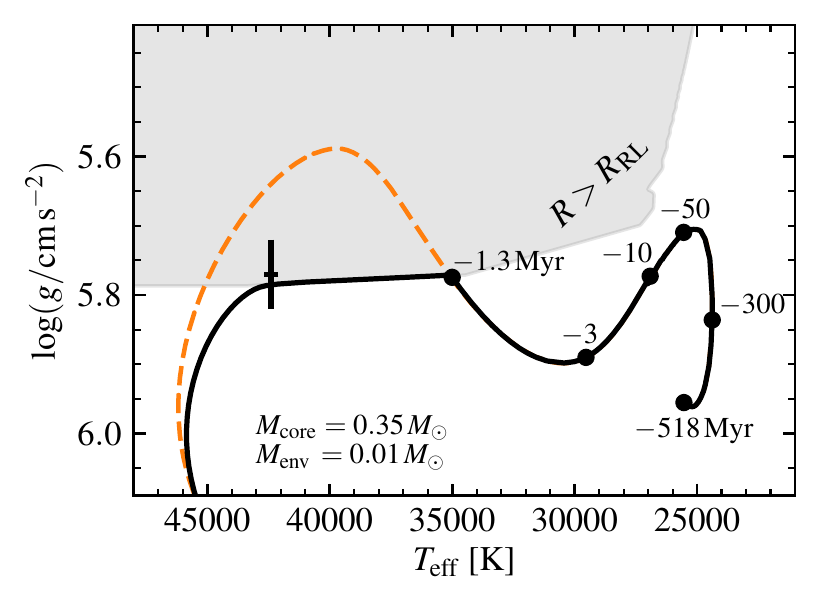}
    \caption{
    {\it Left:} Evolutionary tracks for hot subdwarf models with two values of He mass and envelope mass. The black cross corresponds to the observational constraints given in Table~\ref{tab:system}. Tracks start on the lower right of the plot and evolve leftward toward hotter temperatures when core He is exhausted.
    {\it Right:} Binary evolution track for the most massive of our hot subdwarf models. The gray shaded region shows the maximum radius $R$ that the subdwarf can reach before overflowing its Roche lobe radius $R_{\rm RL}$, which shrinks over time due to gravitational-wave radiation. The black points label ages (in Myr) along the track relative to the present time.
    }
    \label{fig:logg-Teff}
\end{figure*}

Once these models begin core He burning, we remove most of the mass of the outer envelope, leaving only $\approx\,$0.01~$M_\odot$ of H/He envelope material outside the He-dominated core. The star then evolves to become a He-burning hot subdwarf. The left panel of Figure~\ref{fig:logg-Teff} shows evolutionary tracks for a selection of these hot subdwarf models varying both the He core mass and envelope mass. Our {\tt MESA} models for this stage employ the predictive mixing scheme for convection to allow proper growth of the convective core and yield correct He-burning lifetimes \citep{pax18}. These models are relatively low-mass and low-luminosity for hot subdwarfs, and core He burning lasts approximately 500~Myr. This stage corresponds to the portion of the tracks in the lower-temperature ($T_{\rm eff} \leq 30{,}000\ \rm K$) regime in Fig.~\ref{fig:logg-Teff}. Once burning exhausts He in the core, the model evolves toward hotter temperatures over a timescale of order 10~Myr. The core contracts, and residual H in the envelope begins to burn in a shell, pushing the surface to a larger radius and forming the hotter peak in the tracks shown in Fig.~\ref{fig:logg-Teff}. The left panel of Fig.~\ref{fig:logg-Teff} shows that we prefer a mass slightly higher than 0.33~$M_\odot$ to achieve a high enough temperature to match our observations, and a relatively thick H/He envelope to achieve a radius consistent with the measured $\log(g)$.

For the most massive of our four hot subdwarf models, we perform a {\tt MESA} binary evolution calculation with a 0.55~$M_\odot$ WD companion initialized with an orbital period of 148 minutes at the beginning of core He burning. The right panel of Fig.~\ref{fig:logg-Teff} shows the evolutionary track compared to the single-star model and the radius constraint due to its binary Roche lobe (gray shaded region). For the binary evolution, Roche lobe radii are computed using the fit of \citet{Eggleton1983}, and mass transfer rates follow the prescription of \citet{Ritter1988} when the donor star overflows its Roche lobe. In our {\tt MESA} models, the binary separation evolves according to gravitational-wave radiation and angular momentum conservation of material transferred from the donor to its companion \citep{pax15}. We treat the accreting WD as a point mass that accretes all material lost by the Roche lobe overflowing donor star.

In our binary evolution model, the orbit decays due to gravitational-wave losses over the core He-burning lifetime, but not enough to bring the star into contact with its WD companion at this stage. Instead, the core contracts as He is exhausted in the center, and this causes the residual H shell to begin burning. The expansion driven by this shell burning pushes the radius outward to overflow its Roche lobe at an orbital period of 40.5~minutes. The evolution of the envelope drives mass transfer at a rate of $10^{-9}\ M_\odot\, {\rm yr}^{-1}$ lasting approximately 1~Myr as the subdwarf continues to evolve toward hotter temperatures. Accretion onto the WD companion at this rate will cause unstable hydrogen ignition after $\approx 10^{-4}\ M_\odot$ accumulates, leading to a classical novae eruptions \citep{Nomoto82,Nomoto07,Wolf13}. This accretion rate therefore predicts a recurrence time of order $10^5$ years for a total of approximately 10 novae. The novae will cause some mass to be lost from the system, and this effect is not captured in our modeling where we treat the accretor as a point mass that retains all accreted mass. However, the total amount of mass lost by the donor through this phase is only $10^{-3}\ M_\odot$, so the mass transfer efficiency will not have a significant impact on the orbital evolution. %\trmc{Is it worth commenting that in the case of CVs, angular momentum loss caused by novae has been hypothesized to destablise mass transfer and cause merging of systems with normal mass white dwarfs like this one? (papers by Schreiber et al and Nelemans et al) Is this system likely to be inherently more stable than the case of CVs?}

\begin{figure}
    \centering
    \includegraphics{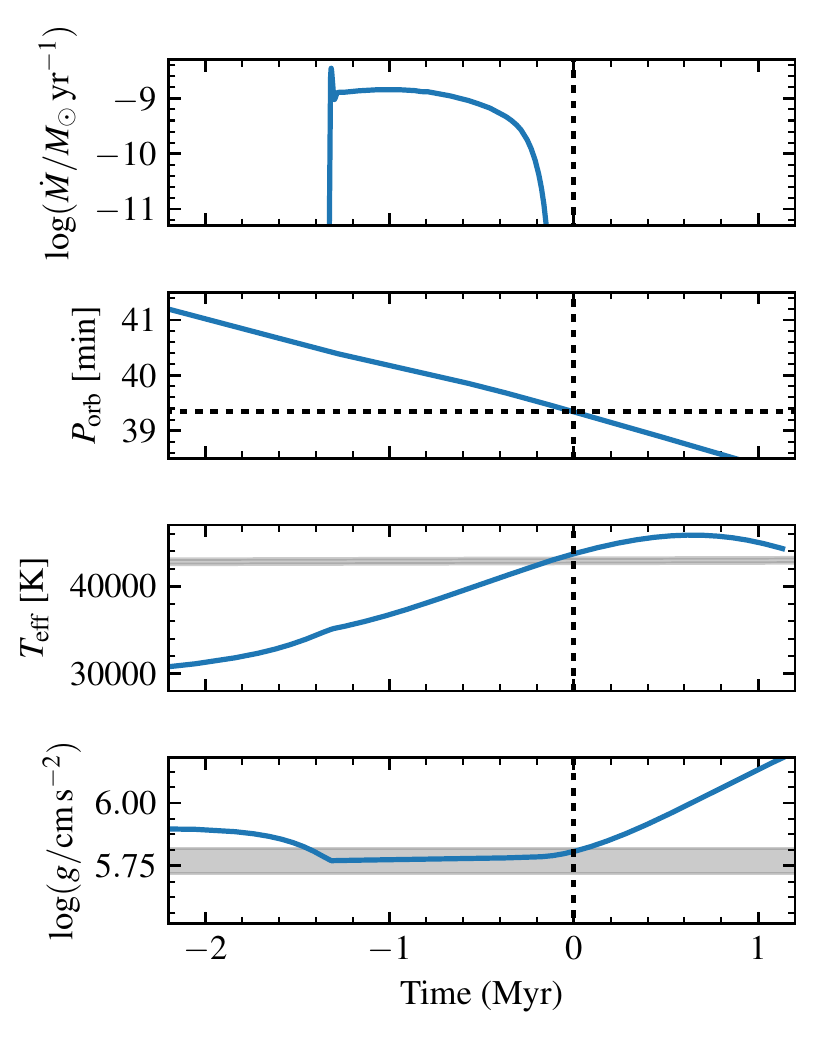}
    \caption{
    Time evolution of the {\tt MESA} binary model through the mass transfer phase leading up to the currently observed state defined by $P_{\rm orb} = 39.34$~minutes at time 0. Gray shaded regions in the lower two panels show the measured values of $T_{\rm eff}$ and $\log(g)$ given in Table~\ref{tab:system}.
    }
    \label{fig:mdot}
\end{figure}

Figure~\ref{fig:mdot} shows the evolution of the donor through this mass transfer phase, which ends just $\approx 10^5$ years before the system reaches the currently observed orbital period and temperature. After 1~Myr of mass transfer, hydrogen burning ceases and the subdwarf contracts to become a WD. While this particular binary evolution model points to a recent cessation of mass transfer in this system, we caution that this may not be a generic feature for the family of hot subdwarf models presented here. The precise configuration of mass transfer in relation to the evolution of $T_{\rm eff}$ and $\log(g)$ is sensitive to both core and envelope mass, and we leave a full exploration of this space of models to other work. We note that in our models there is no He shell burning phase, so the resulting WD retains a substantial ($\approx\,$0.15~$M_\odot$) He layer. While this final He layer mass is subject to theoretical uncertainties in the physics of convective mixing and burning near core He depletion, which affects subsequent He shell structure, models generically predict $\approx$0.1~$M_\odot$ of He will remain after He burning has ceased for subdwarf stars of this mass. This significant mass of He may lead to a thermonuclear supernova in $\approx\,$17~Myr when gravitational waves bring the system into contact again as a double WD binary \citep{Perets19,Zenati19}. If the system does not explode as a thermonuclear supernova, the most likely outcome is a double WD merger and subsequent evolution into an R\,CrB star with a mass of $0.8$-$0.9 M_\odot$, which is the most common mass range for R\,CrB stars \citep{sai08,cla12}. To prevent the merger and form a stable AM\,CVn type system, the system requires a very strong dissipative coupling of the accretor to the orbit and synchronization timescale of $\tau_s \lesssim 0.1$ yr \citep{mar04}.

\subsection{Kinematics of \ztf}
If the sdOB star has evolved from a $2.5$-$2.8$\,\msol\, star, the system has to be part of a young population. To put constraints on the population origin of \ztf\, we calculated its kinematics. The proper motion of the system are taken from the Gaia data release 2 catalog (\citealt{gai18}, $\mu_\alpha$cos$(\delta)=0.009\pm0.047\,\mathrm{mas\,yr^{-1}}$, $\mu_\delta=-1.682\pm0.048\,\mathrm{mas\,yr^{-1}}$). The distance was taken as $1.20\pm0.06$\,kpc as derived from the Gaia parallax (see Sec.\,\ref{sec:system}) and the systemic velocity was taken from the radial velocity curve ($\gamma=-$33.9$\pm$1.9\,\kms, see section\,\ref{orb_atm_pars}).  

\begin{figure*}
    \centering
    \includegraphics[width=0.49\textwidth]{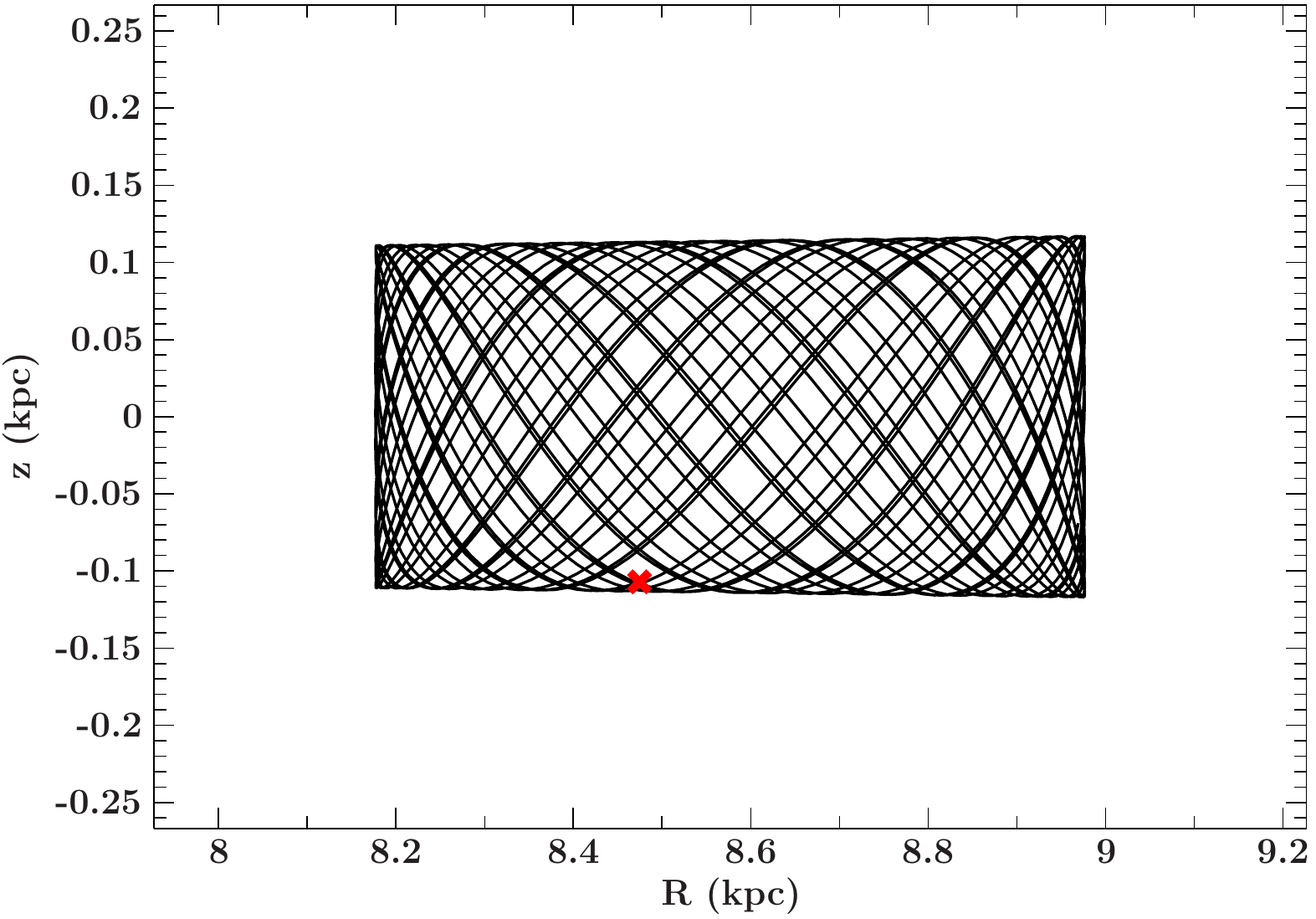} 
    \includegraphics[width=0.47\textwidth]{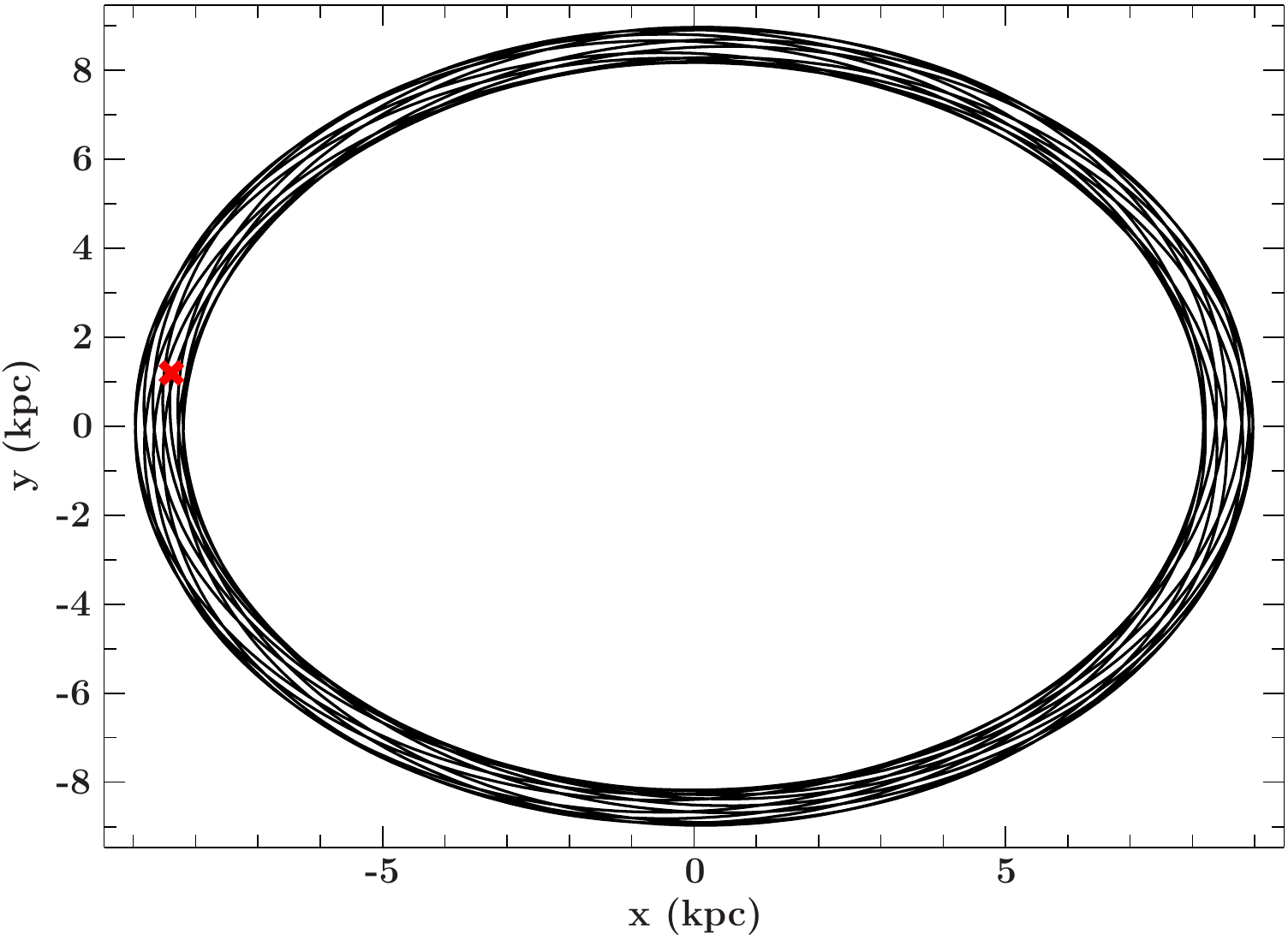}
    \caption{Left: The orbit of \ztf\, projected on to the $x$–$y$ plane. Right: The orbit projected in the $R$, $z$ plane, with $R = \sqrt{x^2+y^2}$-$z$. $x,y$, and $z$ are the Galactic coordinates of the source. The red dot shows the current location of \ztf.}
    \label{fig:kinematics}
\end{figure*}

We employed the approach described in \citet{ode92} and \citet{pau06}. We use the Galactic potential of \citet{all91} as revised by \citet{irr13}. The orbit was integrated from the present to 3 Gyr into the past. The kinematics of \ztf\, are visualized in Fig.\,\ref{fig:kinematics}, where the two panels show the orbit projected on to the $x-y$ and the $R = \sqrt{x^2+y^2}$--$z$ planes, $x,y,z$ being Galactic coordinates. \ztf\, moves within a height of 100\,parsec of the Galactic equator. From the Galactic orbit we conclude that \ztf\, is a member of the Galactic thin disk population.

\subsection{Helium white dwarf interpretation for the donor star}
\cite{kup17a} reported the discovery of a similar sdOB+WD system with an orbital period of 44~minutes and $T_{\rm eff} = 39{,}400\ \rm K$, but no obvious signs of an accretion disk. Their interpretation of that discovery was that the sdOB star in that system is a young He-core WD that is just beginning to cool, and that the binary had exited the common envelope phase within the last Myr. We cannot rule out a similar interpretation for the system presented here, but we prefer the binary evolution models that we present in this work as a more natural explanation for the presence of an accretion disk as a result of recent Roche lobe overflow and mass transfer. We note that an accretion disk around the companion WD cannot be ruled out for the \cite{kup17a} system as the sdOB would outshine an accretion disk and its inclination is too low to show eclipses. Therefore that system may also be consistent with a binary scenario from the family of models we present in this work.

\subsection{White dwarf accretor mass}
We find a mass for the WD companion $M_{\rm WD}=0.545\pm0.020$\,\msol, which is slightly below the typical mass of $0.6$\,\msol. In the standard picture the WD companion was formed first and therefore the main sequence mass of the WD companion had to be larger than the main sequence mass of the sdOB star, which we found earlier to be $2.5$-$2.8$\,\msol. \citet{cum18} present the initial-final mass relation (IFMR) based on 73 WDs for isolated WDs. Using equation 4 from that analysis we find that a main sequence star with $M=2.8$\,\msol\, will form a WD with $M_{\rm WD}=0.71\pm0.09$\,\msol, which is inconsistent with our result.

\citet{kal14} showed that the core-mass and hence the WD mass grows by $\approx10\,\%$ for stars with initial masses $\approx3$\,\msol\, on the asymptotic giant branch (AGB). It is likely that the progenitor of the WD companion experienced a phase of mass transfer on the AGB, where the star lost its envelope before the core has grown to its final mass predicted by the IMFR for isolated WDs and instead formed a WD with a mass $\approx10\,\%$ below the IMFR prediction, and hence ended up with mass consistent with our observed value.

\subsection{An unusual accretion disk}
The accretion disk in \ztf\, is unusual in that it is heavily irradiated by the mass donor. This has one important consequence: even if the accretion rate is significantly lower than we estimate, such that one would normally expect to see dwarf nova outbursts, it could be that the irradiation from the donor suppresses the outbursts by keeping the disk in a permanent high state as has been hypothesized to explain the long outbursts of some X-ray transients \citep{kin98}.

Our high signal-to-noise ratio WHT spectra ($SNR\approx100$) show no evidence for any disk lines. Therefore, we can limit the contribution of the accretion disk to the overall luminosity to $\leq3\,\%$. Our models predict an accretion rate of $10^{-9}$M$_\odot$yr$^{-1}$ or even lower if the system is close to the cessation of its accretion phase. From that we can limit the accretion luminosity to be $<1$\,L$_\odot$ which is significantly smaller than the luminosity of the sdOB star ($L_{\rm{sdOB}}=41\pm9$\,L$_\odot$, see Sec.\,\ref{sec:gaia}) and in agreement with the absence of any signs of the disk in the optical spectra.

\citet{riv19} reported the non-detection of X-rays in a 1\,ks observation with the Neil Gehrels Swift Observatory. X-rays from disk-accreting WDs are generally emitted from the boundary layer. However, with an increasing accretion rate, the boundary layer becomes optically-thick to its own radiation, and emission shifts from the X-ray region to the extreme ultraviolet \citep{pri79, pat85}. Such accretion discs are less luminous in X-ray, despite their higher accretion rates. This has been observed by \citet{whe03}, who conducted a multi-wavelength campaign during a dwarf nova outburst of SS\,Cyg. They showed that during the outburst, as the accretion increases, the X-ray luminosity drops and the extreme UV emission increases. This illustrates the rather complex relation between accretion rate and X-ray flux and perhaps can explain the non-detection of X-rays in \ztf. However, we cannot exclude that our models overestimate the accretion rate, in particular if the sdOB donor is close to the end of mass transfer. We encourage deeper X-ray observations in the future.

%\trmc{Could also discuss: (i) the lack of any line emission (simply too faint perhaps?), (ii) the expected brightness of the disk given the mass transfer rate of Fig.~9. I note that IX~Velorum, a well-known, bright and fairly face-on novalike, has an absolute G mag of 4.5 but an accretion rate of $\approx 5\times 10^{-9}\,$\msol\,yr$^{-1}$ (Linell  et al 2007). Given the lower accretion rate we predict, we can expect the disk here to be fainter, it's also much smaller than the disk in IX Vel, and it's fairly edge-on, so together these effects mean that  it is not too surprising that the light in the system is dominated by the sdOB. We should definitely apply for HST data on this target. The disk may be brighter in the UV.}

\subsection{Gravitational waves}
Due to its short period, \ztf\, is expected to be a strong source of gravitational waves and might be detectable with the Laser Interferometer Space Antenna (\emph{LISA}) as an individual source \citep{ama17}. The gravitational-wave strain amplitude $h$ scales with the masses of both binary components, the binary inclination, the orbital period and the distance of the system.

Based on \citet[and references therein]{kup18} we can calculate the dimensionless gravitational-wave amplitude ($\mathcal{A}$):
\begin{equation}
\mathcal{A} = \frac{2(G\mathcal{M})^{5/3}}{c^4d}(\pi f)^{2/3},
\end{equation}
where $\mathcal{M}$ is the chirp mass, $\mathcal{M} \equiv (M_{\rm sdOB}M_{\rm WD})^{3/5}/(M_{\rm sdOB}+M_{\rm WD})^{1/5}$, $d$ is the distance and $f$ the gravitational-wave frequency with $f=2/P_{\rm orb}$. We find $\mathcal{A}=(8.40\pm0.65)\times10^{-23}$. 

We can then calculate the characteristic strain ($h_c$):
\begin{equation}
h_c = \sqrt{N_{\rm cycle}}\mathcal{A},
\end{equation}
where $N_{\rm cycle}=fT_{\rm obs}$. Assuming the nominal mission lifetime of four years, we find $h_c=(1.96\pm0.14)\times10^{-20}$. For the extended mission lifetime of 10 years, we find  $h_c=(3.10\pm0.22)\times10^{-20}$. Although its high inclination disfavors the signal-to-noise ratio for \emph{LISA}, the presence of an eclipse allows for the determination of the binary parameter to a high accuracy and hence a precise prediction of the gravitational-wave strain. Using the approach described in \citet{bur19} we find a signal-to-noise ratio for \emph{LISA} of $\approx3$ assuming a four year lifetime and $\approx5$ assuming a 10 year lifetime.

Assuming the evolution of the system is governed by gravitational-wave radiation, we can predict the orbital decay of the system $\dot{P}$: 
\begin{equation}
\dot{P}  = \frac{192}{5} \pi^{8/3} \left( \frac{G^{5/3}M_{\rm sdOB}M_{\rm WD}P_{\rm orb}^{-5/3}}{c^5(M_{\rm sdOB}+M_{\rm WD})^{1/3}2^{-5/3}} \right).
\end{equation}
Using this equation we find $\dot{P}=(-1.68\pm0.42)\times10^{-12}$s\,s$^{-1}$. Given the sharp eclipses of the accretion disk, this should be detectable after a few years of monitoring.

\subsection{Selection effect}
The residual hydrogen shell-burning lifetime of the hot subdwarf with $M_{\rm sdOB}=0.35$\,\msol\, in \ztf\, is a factor of $\approx50$-$100$ shorter than its helium core burning lifetime, hence finding \ztf\, at the end of hydrogen shell burning means that there should be at least a few tens of detached low-mass hot subdwarfs with WD companions at \porb$\lesssim2.5$\,hrs. So far, only four helium core-burning hot subdwarfs are known to have a WD companion and have \porb$\lesssim2.5$\,hrs: CD--30$^{\circ}$11223 (\porb$=70.5$\,min; \citealt{gei13, ven12}), PTF1\,J0823+0819 (\porb$=87$\,min; \citealt{kup17}), KPD\,0422+5421 (\porb$=129$\,min; \citealt{koe98,oro99}), KPD\,1930+2752 (\porb$=136$\,min; \citealt{max00, gei07}). Only CD--30$^{\circ}$11223 will start mass transfer to the WD during He-core burning, whereas PTF1\,J0823+0819 might start accreting towards the end of He-shell burning. 

\ztf\, was initially discovered in a search for periodic objects in the \citet{gei19} hot subdwarf catalog of $\approx40,000$ hot subdwarf candidates. Photometric surveys like the ZTF are only sensitive to hot subdwarf binaries with compact companions if the sdB star shows at least a few percent photometric amplitudes from ellipsoidal deformation, hence the sdB has to be close to Roche lobe-filling. As shown in Fig.\,\ref{fig:logg-Teff}, an object like \ztf\, is far from Roche lobe-filling during He-core burning and would not show any photometric variability. 

A different way to detect compact sdB binaries is from large radial-velocity shifts on short timescales. The MUCHFUSS survey used this strategy to find compact hot subdwarf binaries \citep{gei11a} in SDSS multi-epoch spectra and indeed CD--30$^{\circ}$11223 was first identified from large radial-velocity shifts on short timescales. \ztf\, would show a velocity semi-amplitude of $K\approx$250-300\,\kms\, during He-core burning when the binary had \porb$\approx2$\,hrs. This would be easily detectable in low-resolution multi-epoch spectra such as those taken in SDSS with the BOSS spectrograph. However, the hot subdwarf in \ztf\, has evolved from a $2.5\text{--}2.8$\,\msol\, main sequence star and therefore has to be a young system which is consistent with its kinematics. SDSS almost exclusively observed hot subdwarfs at high Galactic latitudes and therefore did not cover the young population of hot subdwarf binaries. Future multi-epoch spectroscopic surveys which cover lower Galactic latitudes will be able to detect a substantial number of progenitor systems of \ztf.

\section{Conclusion and Summary}
\ztf\, was discovered as a short-period variable with a remarkable light curve shape as part of a search for periodic objects in the \citet{gei19} hot subdwarf candidate catalog. Follow-up observations show that \ztf\, is an ultracompact sdOB binary with a compact companion with \porb=$39.3401(1)$\,min making it the most compact hot subdwarf binary known today.

High signal-to-noise ratio photometry obtained with HiPERCAM allow us to put tight constraints on the system parameters. We find that we can only fit the HiPERCAM light curve when including an irradiated accretion disk, making \ztf\, the first known hot subdwarf binary where the sdOB fills its Roche lobe and started mass transfer to its WD companion. Combining the HiPERCAM light curves with spectroscopy, we find a mass ratio $q = M_{\rm sdOB}/M_{\rm WD}=0.617\pm0.015$, a mass for the sdOB  $M_{\rm sdOB}=0.337\pm0.015$\,\msol\,and a WD companion mass $M_{\rm WD}=0.545\pm0.020$\,\msol. The derived sdOB mass is consistent with the estimate from the Gaia parallax and lower than the canonical mass for hot subdwarfs of $\approx0.47$\,\msol. Therefore, the sdOB has not evolved from the standard hot subdwarf channel where the envelope of the hot subdwarf progenitor gets stripped at the tip of the red-giant branch. Instead, it has likely evolved from a $2.5\text{--}2.8$\,\msol\, progenitor which was stripped when crossing the Hertzsprung gap. Therefore, the system has to be young, which is consistent with the observed kinematics.

To put constraints on the evolutionary history of the system, we compared the derived \porb, \teff, \logg\, and mass to evolutionary tracks for He-stars computed with \texttt{MESA}. We find that the binary left the common envelope when the hot subdwarf was formed at \porb$\approx150$\,min and reached contact at \porb$\approx40$\,min during residual hydrogen shell burning when the envelope started to expand again. We currently observe the object towards the end of hydrogen shell burning. Once hydrogen shell burning is finished, the sdOB will shrink within its Roche lobe and the binary will reach contact again after $\approx\,$17~Myr as a double WD. The system will either explode as a thermonuclear supernova \citep{Perets19,Zenati19} or form an R\,CrB star. 

Although the He-core burning lifetime is a factor of $\approx 100$ larger compared to the residual hydrogen shell burning lifetime, current surveys are not sensitive enough to detect such systems when they are part of the young stellar population and far away from Roche lobe-filling. Ongoing and upcoming multi-epoch spectroscopic surveys which cover low Galactic latitudes, like LAMOST \citep{den12}, SDSS-V \citep{kol17}, WEAVE \citep{dal12} or 4MOST \citep{deJ19} will be sensitive to explore the young population of detached sdB binaries with compact companions and periods $\lesssim$\,few hours.

\section{Acknowledgments}

Based on observations obtained with the Samuel Oschin Telescope 48-inch at the Palomar Observatory as part of the Zwicky Transient Facility project. ZTF is supported by the National Science Foundation under Grant No. AST-1440341 and a collaboration including Caltech, IPAC, the Weizmann Institute for Science, the Oskar Klein Center at Stockholm University, the University of Maryland, the University of Washington, Deutsches Elektronen-Synchrotron and Humboldt University, Los Alamos National Laboratories, the TANGO Consortium of Taiwan, the University of Wisconsin at Milwaukee, and Lawrence Berkeley National Laboratories. Operations are conducted by COO, IPAC, and UW.

Some of the data presented herein were obtained at the W.M. Keck Observatory, which is operated as a scientific partnership among the California Institute of Technology, the University of California and the National Aeronautics and Space Administration. The Observatory was made possible by the generous financial support of the W.M. Keck Foundation. The authors wish to recognize and acknowledge the very significant cultural role and reverence that the summit of Mauna Kea has always had within the indigenous Hawaiian community. We are most fortunate to have the opportunity to conduct observations from this mountain.

Some results presented in this paper are based on observations made with the WHT operated on the island of La Palma by the Isaac Newton Group in the Spanish Observatorio del Roque de los Muchachos of the Institutio de Astrofisica de Canarias.

Based on observations made with the Gran Telescopio Canarias (GTC), installed at the Spanish Observatorio del Roque de los Muchachos of the Instituto de Astrofísica de Canarias, in the island of La Palma.

The KPED team thanks the National Science Foundation and the National Optical Astronomical Observatory for making the Kitt Peak 2.1-m telescope available. The KPED team thanks the National Science Foundation, the National Optical Astronomical Observatory and the Murty family for support in the building and operation of KPED. 

This research was supported in part by the National Science Foundation through grant ACI-1663688, and at the KITP by grant PHY-1748958. This research benefited from interactions that were funded by the Gordon and Betty Moore Foundation through Grant GBMF5076. 

We acknowledge the use of the Center for Scientific Computing supported by the California NanoSystems Institute and the Materials Research Science and Engineering Center (MRSEC) at UC Santa Barbara through NSF DMR 1720256 and NSF CNS 1725797.

HiPERCAM and VSD are funded by the European Research Council under the European Union’s Seventh Framework Programme (FP/2007-2013) under ERC-2013-ADG Grant Agreement no. 340040 (HiPERCAM).

We thank Brad Barlow, Josiah Schwab and Stephan Geier for helpful conversations. MC is supported by the David and Ellen Lee Postdoctoral Fellowship at the California Institute of Technology. TRM was supported by a grant from the United Kingdom's Science and Technology Facilities Council. PS acknowledges support from NSF grant AST-1514737. 

This work has made use of data from the European Space Agency (ESA) mission {\it Gaia} (\url{https://www.cosmos.esa.int/gaia}), processed by the {\it Gaia} Data Processing and Analysis Consortium (DPAC,
\url{https://www.cosmos.esa.int/web/gaia/dpac/consortium}). Funding for the DPAC has been provided by national institutions, in particular the institutions
participating in the {\it Gaia} Multilateral Agreement.

This work benefited from a workshop held at DARK in July 2019 that was funded by the Danish National Research Foundation (DNRF132).
We thank Josiah Schwab for his efforts in organising this.

\vspace{5mm}
\facilities{PO:1.2m (ZTF), Hale (DBSP), Keck:I (LRIS), ING:Herschel (ISIS), GTC (HiPERCAM)}

\software{\texttt{Lpipe} \citep{per19}, \texttt{PyRAF} \citep{bel16}, \texttt{Gatspy} \citep{van15, van15a}, \texttt{FITSB2} \citep{nap04a}, \texttt{LCURVE} \citep{cop10}, \texttt{emcee} \citep{for13}, \texttt{MESA} \citep{pax11,pax13,pax15,pax18,pax19}, \texttt{Matplotlib} \citep{hun07}, \texttt{Astropy} \citep{astpy13, astpy18}, \texttt{Numpy} \citep{numpy}}

%\software{}

\bibliographystyle{aasjournal}
\bibliography{refs,refs_1508}

\begin{thebibliography}{}
\expandafter\ifx\csname natexlab\endcsname\relax\def\natexlab#1{#1}\fi
\providecommand{\url}[1]{\href{#1}{#1}}

\bibitem[{{Allen} \& {Santillan}(1991)}]{all91}
{Allen}, C., \& {Santillan}, A. 1991, \rmxaa, 22, 255

\bibitem[{{Amaro-Seoane} {et~al.}(2017){Amaro-Seoane}, {Audley}, {Babak},
  {Baker}, {Barausse}, {Bender}, {Berti}, {Binetruy}, {Born}, {Bortoluzzi},
  {Camp}, {Caprini}, {Cardoso}, {Colpi}, {Conklin}, {Cornish}, {Cutler},
  {Danzmann}, {Dolesi}, {Ferraioli}, {Ferroni}, {Fitzsimons}, {Gair}, {Gesa
  Bote}, {Giardini}, {Gibert}, {Grimani}, {Halloin}, {Heinzel}, {Hertog},
  {Hewitson}, {Holley-Bockelmann}, {Hollington}, {Hueller}, {Inchauspe},
  {Jetzer}, {Karnesis}, {Killow}, {Klein}, {Klipstein}, {Korsakova}, {Larson},
  {Livas}, {Lloro}, {Man}, {Mance}, {Martino}, {Mateos}, {McKenzie},
  {McWilliams}, {Miller}, {Mueller}, {Nardini}, {Nelemans}, {Nofrarias},
  {Petiteau}, {Pivato}, {Plagnol}, {Porter}, {Reiche}, {Robertson},
  {Robertson}, {Rossi}, {Russano}, {Schutz}, {Sesana}, {Shoemaker}, {Slutsky},
  {Sopuerta}, {Sumner}, {Tamanini}, {Thorpe}, {Troebs}, {Vallisneri},
  {Vecchio}, {Vetrugno}, {Vitale}, {Volonteri}, {Wanner}, {Ward}, {Wass},
  {Weber}, {Ziemer}, \& {Zweifel}}]{ama17}
{Amaro-Seoane}, P., {Audley}, H., {Babak}, S., {et~al.} 2017, ArXiv e-prints,
  arXiv:1702.00786

\bibitem[{{Astropy Collaboration} {et~al.}(2013){Astropy Collaboration},
  {Robitaille}, {Tollerud}, {Greenfield}, {Droettboom}, {Bray}, {Aldcroft},
  {Davis}, {Ginsburg}, {Price-Whelan}, {Kerzendorf}, {Conley}, {Crighton},
  {Barbary}, {Muna}, {Ferguson}, {Grollier}, {Parikh}, {Nair}, {Unther},
  {Deil}, {Woillez}, {Conseil}, {Kramer}, {Turner}, {Singer}, {Fox}, {Weaver},
  {Zabalza}, {Edwards}, {Azalee Bostroem}, {Burke}, {Casey}, {Crawford},
  {Dencheva}, {Ely}, {Jenness}, {Labrie}, {Lim}, {Pierfederici}, {Pontzen},
  {Ptak}, {Refsdal}, {Servillat}, \& {Streicher}}]{astpy13}
{Astropy Collaboration}, {Robitaille}, T.~P., {Tollerud}, E.~J., {et~al.} 2013,
  \aap, 558, A33

\bibitem[{{Astropy Collaboration} {et~al.}(2018){Astropy Collaboration},
  {Price-Whelan}, {Sip{\H o}cz}, {G{\"u}nther}, {Lim}, {Crawford}, {Conseil},
  {Shupe}, {Craig}, {Dencheva}, {Ginsburg}, {VanderPlas}, {Bradley},
  {P{\'e}rez-Su{\'a}rez}, {de Val-Borro}, {Aldcroft}, {Cruz}, {Robitaille},
  {Tollerud}, {Ardelean}, {Babej}, {Bach}, {Bachetti}, {Bakanov}, {Bamford},
  {Barentsen}, {Barmby}, {Baumbach}, {Berry}, {Biscani}, {Boquien}, {Bostroem},
  {Bouma}, {Brammer}, {Bray}, {Breytenbach}, {Buddelmeijer}, {Burke},
  {Calderone}, {Cano Rodr{\'{\i}}guez}, {Cara}, {Cardoso}, {Cheedella},
  {Copin}, {Corrales}, {Crichton}, {D'Avella}, {Deil}, {Depagne}, {Dietrich},
  {Donath}, {Droettboom}, {Earl}, {Erben}, {Fabbro}, {Ferreira}, {Finethy},
  {Fox}, {Garrison}, {Gibbons}, {Goldstein}, {Gommers}, {Greco}, {Greenfield},
  {Groener}, {Grollier}, {Hagen}, {Hirst}, {Homeier}, {Horton}, {Hosseinzadeh},
  {Hu}, {Hunkeler}, {Ivezi{\'c}}, {Jain}, {Jenness}, {Kanarek}, {Kendrew},
  {Kern}, {Kerzendorf}, {Khvalko}, {King}, {Kirkby}, {Kulkarni}, {Kumar},
  {Lee}, {Lenz}, {Littlefair}, {Ma}, {Macleod}, {Mastropietro}, {McCully},
  {Montagnac}, {Morris}, {Mueller}, {Mumford}, {Muna}, {Murphy}, {Nelson},
  {Nguyen}, {Ninan}, {N{\"o}the}, {Ogaz}, {Oh}, {Parejko}, {Parley}, {Pascual},
  {Patil}, {Patil}, {Plunkett}, {Prochaska}, {Rastogi}, {Reddy Janga},
  {Sabater}, {Sakurikar}, {Seifert}, {Sherbert}, {Sherwood-Taylor}, {Shih},
  {Sick}, {Silbiger}, {Singanamalla}, {Singer}, {Sladen}, {Sooley},
  {Sornarajah}, {Streicher}, {Teuben}, {Thomas}, {Tremblay}, {Turner},
  {Terr{\'o}n}, {van Kerkwijk}, {de la Vega}, {Watkins}, {Weaver}, {Whitmore},
  {Woillez}, {Zabalza}, \& {Astropy Contributors}}]{astpy18}
{Astropy Collaboration}, {Price-Whelan}, A.~M., {Sip{\H o}cz}, B.~M., {et~al.}
  2018, \aj, 156, 123

\bibitem[{{Bauer} {et~al.}(2017){Bauer}, {Schwab}, \& {Bildsten}}]{bau17}
{Bauer}, E.~B., {Schwab}, J., \& {Bildsten}, L. 2017, \apj, 845, 97

\bibitem[{{Bellm} \& {Sesar}(2016)}]{bel16}
{Bellm}, E.~C., \& {Sesar}, B. 2016, {pyraf-dbsp: Reduction pipeline for the
  Palomar Double Beam Spectrograph}, Astrophysics Source Code Library, , ,
  ascl:1602.002

\bibitem[{{Bellm} {et~al.}(2019{\natexlab{a}}){Bellm}, {Kulkarni}, {Graham},
  {Dekany}, {Smith}, {Riddle}, {Masci}, {Helou}, {Prince}, {Adams},
  {Barbarino}, {Barlow}, {Bauer}, {Beck}, {Belicki}, {Biswas}, {Blagorodnova},
  {Bodewits}, {Bolin}, {Brinnel}, {Brooke}, {Bue}, {Bulla}, {Burruss}, {Cenko},
  {Chang}, {Connolly}, {Coughlin}, {Cromer}, {Cunningham}, {De}, {Delacroix},
  {Desai}, {Duev}, {Eadie}, {Farnham}, {Feeney}, {Feindt}, {Flynn},
  {Franckowiak}, {Frederick}, {Fremling}, {Gal-Yam}, {Gezari}, {Giomi},
  {Goldstein}, {Golkhou}, {Goobar}, {Groom}, {Hacopians}, {Hale}, {Henning},
  {Ho}, {Hover}, {Howell}, {Hung}, {Huppenkothen}, {Imel}, {Ip}, {Ivezi{\'c}},
  {Jackson}, {Jones}, {Juric}, {Kasliwal}, {Kaspi}, {Kaye}, {Kelley},
  {Kowalski}, {Kramer}, {Kupfer}, {Landry}, {Laher}, {Lee}, {Lin}, {Lin},
  {Lunnan}, {Giomi}, {Mahabal}, {Mao}, {Miller}, {Monkewitz}, {Murphy},
  {Ngeow}, {Nordin}, {Nugent}, {Ofek}, {Patterson}, {Penprase}, {Porter},
  {Rauch}, {Rebbapragada}, {Reiley}, {Rigault}, {Rodriguez}, {van Roestel},
  {Rusholme}, {van Santen}, {Schulze}, {Shupe}, {Singer}, {Soumagnac}, {Stein},
  {Surace}, {Sollerman}, {Szkody}, {Taddia}, {Terek}, {Van Sistine}, {van
  Velzen}, {Vestrand}, {Walters}, {Ward}, {Ye}, {Yu}, {Yan}, \&
  {Zolkower}}]{bel19}
{Bellm}, E.~C., {Kulkarni}, S.~R., {Graham}, M.~J., {et~al.}
  2019{\natexlab{a}}, \pasp, 131, 018002

\bibitem[{{Bellm} {et~al.}(2019{\natexlab{b}}){Bellm}, {Kulkarni}, {Barlow},
  {Feindt}, {Graham}, {Goobar}, {Kupfer}, {Ngeow}, {Nugent}, {Ofek}, {Prince},
  {Riddle}, {Walters}, \& {Ye}}]{bel19a}
{Bellm}, E.~C., {Kulkarni}, S.~R., {Barlow}, T., {et~al.} 2019{\natexlab{b}},
  \pasp, 131, 068003

\bibitem[{{Bildsten} {et~al.}(2007){Bildsten}, {Shen}, {Weinberg}, \&
  {Nelemans}}]{bil07}
{Bildsten}, L., {Shen}, K.~J., {Weinberg}, N.~N., \& {Nelemans}, G. 2007, ApJL,
  662, L95

\bibitem[{{Bloemen} {et~al.}(2011){Bloemen}, {Marsh}, {{\O}stensen},
  {Charpinet}, {Fontaine}, {Degroote}, {Heber}, {Kawaler}, {Aerts}, {Green},
  {Telting}, {Brassard}, {G{\"a}nsicke}, {Handler}, {Kurtz}, {Silvotti}, {Van
  Grootel}, {Lindberg}, {Pursimo}, {Wilson}, {Gilliland}, {Kjeldsen},
  {Christensen-Dalsgaard}, {Borucki}, {Koch}, {Jenkins}, \& {Klaus}}]{blo11}
{Bloemen}, S., {Marsh}, T.~R., {{\O}stensen}, R.~H., {et~al.} 2011, MNRAS, 410,
  1787

\bibitem[{{Brooks} {et~al.}(2015){Brooks}, {Bildsten}, {Marchant}, \&
  {Paxton}}]{bro15}
{Brooks}, J., {Bildsten}, L., {Marchant}, P., \& {Paxton}, B. 2015, \apj, 807,
  74

\bibitem[{{Brooks} {et~al.}(2017){Brooks}, {Kupfer}, \& {Bildsten}}]{bro17}
{Brooks}, J., {Kupfer}, T., \& {Bildsten}, L. 2017, \apj, 847, 78

\bibitem[{{Burdge} {et~al.}(2019){Burdge}, {Coughlin}, {Fuller}, {Kupfer},
  {Bellm}, {Bildsten}, {Graham}, {Kaplan}, {Roestel}, {Dekany}, {Duev},
  {Feeney}, {Giomi}, {Helou}, {Kaye}, {Laher}, {Mahabal}, {Masci}, {Riddle},
  {Shupe}, {Soumagnac}, {Smith}, {Szkody}, {Walters}, {Kulkarni}, \&
  {Prince}}]{bur19}
{Burdge}, K.~B., {Coughlin}, M.~W., {Fuller}, J., {et~al.} 2019, \nat, 571, 528

\bibitem[{{Carter} {et~al.}(1993)}]{car93}
{Carter}, D., {et~al.} 1993

\bibitem[{{Chambers} {et~al.}(2016){Chambers}, {Magnier}, {Metcalfe},
  {Flewelling}, {Huber}, {Waters}, {Denneau}, {Draper}, {Farrow}, {Finkbeiner},
  {Holmberg}, {Koppenhoefer}, {Price}, {Saglia}, {Schlafly}, {Smartt},
  {Sweeney}, {Wainscoat}, {Burgett}, {Grav}, {Heasley}, {Hodapp}, {Jedicke},
  {Kaiser}, {Kudritzki}, {Luppino}, {Lupton}, {Monet}, {Morgan}, {Onaka},
  {Stubbs}, {Tonry}, {Banados}, {Bell}, {Bender}, {Bernard}, {Botticella},
  {Casertano}, {Chastel}, {Chen}, {Chen}, {Cole}, {Deacon}, {Frenk},
  {Fitzsimmons}, {Gezari}, {Goessl}, {Goggia}, {Goldman}, {Grebel}, {Hambly},
  {Hasinger}, {Heavens}, {Heckman}, {Henderson}, {Henning}, {Holman}, {Hopp},
  {Ip}, {Isani}, {Keyes}, {Koekemoer}, {Kotak}, {Long}, {Lucey}, {Liu},
  {Martin}, {McLean}, {Morganson}, {Murphy}, {Nieto-Santisteban}, {Norberg},
  {Peacock}, {Pier}, {Postman}, {Primak}, {Rae}, {Rest}, {Riess}, {Riffeser},
  {Rix}, {Roser}, {Schilbach}, {Schultz}, {Scolnic}, {Szalay}, {Seitz},
  {Shiao}, {Small}, {Smith}, {Soderblom}, {Taylor}, {Thakar}, {Thiel},
  {Thilker}, {Urata}, {Valenti}, {Walter}, {Watters}, {Werner}, {White},
  {Wood-Vasey}, \& {Wyse}}]{cham16}
{Chambers}, K.~C., {Magnier}, E.~A., {Metcalfe}, N., {et~al.} 2016, arXiv
  e-prints, arXiv:1612.05560

\bibitem[{{Choi} {et~al.}(2016){Choi}, {Dotter}, {Conroy}, {Cantiello},
  {Paxton}, \& {Johnson}}]{cho16}
{Choi}, J., {Dotter}, A., {Conroy}, C., {et~al.} 2016, \apj, 823, 102

\bibitem[{{Claret} \& {Bloemen}(2011)}]{cla11}
{Claret}, A., \& {Bloemen}, S. 2011, \aap, 529, A75

\bibitem[{{Clayton}(2012)}]{cla12}
{Clayton}, G.~C. 2012, Journal of the American Association of Variable Star
  Observers (JAAVSO), 40, 539

\bibitem[{{Copperwheat} {et~al.}(2010){Copperwheat}, {Marsh}, {Dhillon},
  {Littlefair}, {Hickman}, {G{\"a}nsicke}, \& {Southworth}}]{cop10}
{Copperwheat}, C.~M., {Marsh}, T.~R., {Dhillon}, V.~S., {et~al.} 2010, MNRAS,
  402, 1824

\bibitem[{{Coughlin} {et~al.}(2019){Coughlin}, {Dekany}, {Duev}, {Feeney},
  {Kulkarni}, {Riddle}, {Ahumada}, {Burdge}, {Dugas}, {Fremling}, {Hallinan},
  {Prince}, \& {van Roestel}}]{cou19}
{Coughlin}, M.~W., {Dekany}, R.~G., {Duev}, D.~A., {et~al.} 2019, \mnras, 485,
  1412

\bibitem[{{Cummings} {et~al.}(2018){Cummings}, {Kalirai}, {Tremblay},
  {Ramirez-Ruiz}, \& {Choi}}]{cum18}
{Cummings}, J.~D., {Kalirai}, J.~S., {Tremblay}, P.~E., {Ramirez-Ruiz}, E., \&
  {Choi}, J. 2018, \apj, 866, 21

\bibitem[{{Dalton} {et~al.}(2012){Dalton}, {Trager}, {Abrams}, {Carter},
  {Bonifacio}, {Aguerri}, {MacIntosh}, {Evans}, {Lewis}, {Navarro}, {Agocs},
  {Dee}, {Rousset}, {Tosh}, {Middleton}, {Pragt}, {Terrett}, {Brock}, {Benn},
  {Verheijen}, {Cano Infantes}, {Bevil}, {Steele}, {Mottram}, {Bates},
  {Gribbin}, {Rey}, {Rodriguez}, {Delgado}, {Guinouard}, {Walton}, {Irwin},
  {Jagourel}, {Stuik}, {Gerlofsma}, {Roelfsma}, {Skillen}, {Ridings},
  {Balcells}, {Daban}, {Gouvret}, {Venema}, \& {Girard}}]{dal12}
{Dalton}, G., {Trager}, S.~C., {Abrams}, D.~C., {et~al.} 2012, Society of
  Photo-Optical Instrumentation Engineers (SPIE) Conference Series, Vol. 8446,
  {WEAVE: the next generation wide-field spectroscopy facility for the William
  Herschel Telescope}, 84460P

\bibitem[{{De} {et~al.}(2019){De}, {Kasliwal}, {Polin}, {Nugent}, {Bildsten},
  {Adams}, {Bellm}, {Blagorodnova}, {Burdge}, \& {Cannella}}]{de19}
{De}, K., {Kasliwal}, M.~M., {Polin}, A., {et~al.} 2019, \apj, 873, L18

\bibitem[{{de Jong} {et~al.}(2019){de Jong}, {Agertz}, {Berbel}, {Aird},
  {Alexander}, {Amarsi}, {Anders}, {Andrae}, {Ansarinejad}, {Ansorge},
  {Antilogus}, {Anwand -Heerwart}, {Arentsen}, {Arnadottir}, {Asplund},
  {Auger}, {Azais}, {Baade}, {Baker}, {Baker}, {Balbinot}, {Baldry}, {Banerji},
  {Barden}, {Barklem}, {Barth{\'e}l{\'e}my-Mazot}, {Battistini}, {Bauer},
  {Bell}, {Bellido-Tirado}, {Bellstedt}, {Belokurov}, {Bensby}, {Bergemann},
  {Bestenlehner}, {Bielby}, {Bilicki}, {Blake}, {Bland-Hawthorn}, {Boeche},
  {Boland}, {Boller}, {Bongard}, {Bongiorno}, {Bonifacio}, {Boudon}, {Brooks},
  {Brown}, {Brown}, {Br{\"u}ggen}, {Brynnel}, {Brzeski}, {Buchert},
  {Buschkamp}, {Caffau}, {Caillier}, {Carrick}, {Casagrande}, {Case}, {Casey},
  {Cesarini}, {Cescutti}, {Chapuis}, {Chiappini}, {Childress}, {Christlieb},
  {Church}, {Cioni}, {Cluver}, {Colless}, {Collett}, {Comparat}, {Cooper},
  {Couch}, {Courbin}, {Croom}, {Croton}, {Daguis{\'e}}, {Dalton}, {Davies},
  {Davis}, {de Laverny}, {Deason}, {Dionies}, {Disseau}, {Doel}, {D{\"o}scher},
  {Driver}, {Dwelly}, {Eckert}, {Edge}, {Edvardsson}, {Youssoufi}, {Elhaddad},
  {Enke}, {Erfanianfar}, {Farrell}, {Fechner}, {Feiz}, {Feltzing}, {Ferreras},
  {Feuerstein}, {Feuillet}, {Finoguenov}, {Ford}, {Fotopoulou}, {Fouesneau},
  {Frenk}, {Frey}, {Gaessler}, {Geier}, {Fusillo}, {Gerhard}, {Giannantonio},
  {Giannone}, {Gibson}, {Gillingham}, {Gonz{\'a}lez-Fern{\'a}ndez},
  {Gonzalez-Solares}, {Gottloeber}, {Gould}, {Grebel}, {Gueguen}, {Guiglion},
  {Haehnelt}, {Hahn}, {Hansen}, {Hartman}, {Hauptner}, {Hawkins}, {Haynes},
  {Haynes}, {Heiter}, {Helmi}, {Aguayo}, {Hewett}, {Hinton}, {Hobbs}, {Hoenig},
  {Hofman}, {Hook}, {Hopgood}, {Hopkins}, {Hourihane}, {Howes}, {Howlett},
  {Huet}, {Irwin}, {Iwert}, {Jablonka}, {Jahn}, {Jahnke}, {Jarno}, {Jin},
  {Jofre}, {Johl}, {Jones}, {J{\"o}nsson}, {Jordan}, {Karovicova}, {Khalatyan},
  {Kelz}, {Kennicutt}, {King}, {Kitaura}, {Klar}, {Klauser}, {Kneib}, {Koch},
  {Koposov}, {Kordopatis}, {Korn}, {Kosmalski}, {Kotak}, {Kovalev}, {Kreckel},
  {Kripak}, {Krumpe}, {Kuijken}, {Kunder}, {Kushniruk}, {Lam}, {Lamer},
  {Laurent}, {Lawrence}, {Lehmitz}, {Lemasle}, {Lewis}, {Li}, {Lidman}, {Lind},
  {Liske}, {Lizon}, {Loveday}, {Ludwig}, {McDermid}, {Maguire}, {Mainieri},
  {Mali}, {Mandel}, {Mandel}, {Mannering}, {Martell}, {Martinez Delgado},
  {Matijevic}, {McGregor}, {McMahon}, {McMillan}, {Mena}, {Merloni}, {Meyer},
  {Michel}, {Micheva}, {Migniau}, {Minchev}, {Monari}, {Muller}, {Murphy},
  {Muthukrishna}, {Nandra}, {Navarro}, {Ness}, {Nichani}, {Nichol}, {Nicklas},
  {Niederhofer}, {Norberg}, {Obreschkow}, {Oliver}, {Owers}, {Pai},
  {Pankratow}, {Parkinson}, {Paschke}, {Paterson}, {Pecontal}, {Parry},
  {Phillips}, {Pillepich}, {Pinard}, {Pirard}, {Piskunov}, {Plank},
  {Pl{\"u}schke}, {Pons}, {Popesso}, {Power}, {Pragt}, {Pramskiy}, {Pryer},
  {Quattri}, {Queiroz}, {Quirrenbach}, {Rahurkar}, {Raichoor}, {Ramstedt},
  {Rau}, {Recio-Blanco}, {Reiss}, {Renaud}, {Revaz}, {Rhode}, {Richard},
  {Richter}, {Rix}, {Robotham}, {Roelfsema}, {Romaniello}, {Rosario},
  {Rothmaier}, {Roukema}, {Ruchti}, {Rupprecht}, {Rybizki}, {Ryde}, {Saar},
  {Sadler}, {Sahl{\'e}n}, {Salvato}, {Sassolas}, {Saunders}, {Saviauk},
  {Sbordone}, {Schmidt}, {Schnurr}, {Scholz}, {Schwope}, {Seifert}, {Shanks},
  {Sheinis}, {Sivov}, {Sk{\'u}lad{\'o}ttir}, {Smartt}, {Smedley}, {Smith},
  {Smith}, {Sorce}, {Spitler}, {Starkenburg}, {Steinmetz}, {Stilz}, {Storm},
  {Sullivan}, {Sutherland}, {Swann}, {Tamone}, {Taylor}, {Teillon}, {Tempel},
  {ter Horst}, {Thi}, {Tolstoy}, {Trager}, {Traven}, {Tremblay}, {Tresse},
  {Valentini}, {van de Weygaert}, {van den Ancker}, {Veljanoski}, {Venkatesan},
  {Wagner}, {Wagner}, {Walcher}, {Waller}, {Walton}, {Wang}, {Winkler},
  {Wisotzki}, {Worley}, {Worseck}, {Xiang}, {Xu}, {Yong}, {Zhao}, {Zheng},
  {Zscheyge}, \& {Zucker}}]{deJ19}
{de Jong}, R.~S., {Agertz}, O., {Berbel}, A.~A., {et~al.} 2019, The Messenger,
  175, 3

\bibitem[{{Deng} {et~al.}(2012){Deng}, {Newberg}, {Liu}, {Carlin}, {Beers},
  {Chen}, {Chen}, {Christlieb}, {Grillmair}, {Guhathakurta}, {Han}, {Hou},
  {Lee}, {L{\'e}pine}, {Li}, {Liu}, {Pan}, {Sellwood}, {Wang}, {Wang}, {Yang},
  {Yanny}, {Zhang}, {Zhang}, {Zheng}, \& {Zhu}}]{den12}
{Deng}, L.-C., {Newberg}, H.~J., {Liu}, C., {et~al.} 2012, Research in
  Astronomy and Astrophysics, 12, 735

\bibitem[{{Dhillon} {et~al.}(2018){Dhillon}, {Dixon}, {Gamble}, {Kerry},
  {Littlefair}, {Parsons}, {Marsh}, {Bezawada}, {Black}, {Gao}, {Henry},
  {Lunney}, {Miller}, {Dubbeldam}, {Morris}, {Osborn}, {Wilson}, {Casares},
  {Mu{\~n}oz-Darias}, {Pall{\'e}}, {Rodriguez-Gil}, {Shahbaz}, \& {de Ugarte
  Postigo}}]{dhi18}
{Dhillon}, V., {Dixon}, S., {Gamble}, T., {et~al.} 2018, in Society of
  Photo-Optical Instrumentation Engineers (SPIE) Conference Series, Vol. 10702,
  \procspie, 107020L

\bibitem[{{Dhillon} {et~al.}(2016){Dhillon}, {Marsh}, {Bezawada}, {Black},
  {Dixon}, {Gamble}, {Henry}, {Kerry}, {Littlefair}, {Lunney}, {Morris},
  {Osborn}, \& {Wilson}}]{dhi16}
{Dhillon}, V.~S., {Marsh}, T.~R., {Bezawada}, N., {et~al.} 2016, Society of
  Photo-Optical Instrumentation Engineers (SPIE) Conference Series, Vol. 9908,
  {HiPERCAM: a high-speed quintuple-beam CCD camera for the study of rapid
  variability in the universe}, 99080Y

\bibitem[{{Dotter}(2016)}]{dot16}
{Dotter}, A. 2016, \apjs, 222, 8

\bibitem[{{Dufton}(1972)}]{duf72}
{Dufton}, P.~L. 1972, \mnras, 159, 79

\bibitem[{{Eggleton}(1983)}]{Eggleton1983}
{Eggleton}, P.~P. 1983, \apj, 268, 368

\bibitem[{{Fink} {et~al.}(2010){Fink}, {R{\"o}pke}, {Hillebrandt},
  {Seitenzahl}, {Sim}, \& {Kromer}}]{fin10}
{Fink}, M., {R{\"o}pke}, F.~K., {Hillebrandt}, W., {et~al.} 2010, \aap, 514,
  A53

\bibitem[{{Foreman-Mackey} {et~al.}(2013){Foreman-Mackey}, {Hogg}, {Lang}, \&
  {Goodman}}]{for13}
{Foreman-Mackey}, D., {Hogg}, D.~W., {Lang}, D., \& {Goodman}, J. 2013, \pasp,
  125, 306

\bibitem[{{Gaia Collaboration} {et~al.}(2016){Gaia Collaboration}, {Prusti},
  {de Bruijne}, {Brown}, {Vallenari}, {Babusiaux}, {Bailer-Jones}, {Bastian},
  {Biermann}, {Evans}, \& et~al.}]{gai16}
{Gaia Collaboration}, {Prusti}, T., {de Bruijne}, J.~H.~J., {et~al.} 2016,
  \aap, 595, A1

\bibitem[{{Gaia Collaboration} {et~al.}(2018){Gaia Collaboration}, {Brown},
  {Vallenari}, {Prusti}, {de Bruijne}, {Babusiaux}, {Bailer-Jones}, {Biermann},
  {Evans}, {Eyer}, {Jansen}, {Jordi}, {Klioner}, {Lammers}, {Lindegren},
  {Luri}, {Mignard}, {Panem}, {Pourbaix}, {Randich}, {Sartoretti}, {Siddiqui},
  {Soubiran}, {van Leeuwen}, {Walton}, {Arenou}, {Bastian}, {Cropper},
  {Drimmel}, {Katz}, {Lattanzi}, {Bakker}, {Cacciari}, {Casta{\~n}eda},
  {Chaoul}, {Cheek}, {De Angeli}, {Fabricius}, {Guerra}, {Holl}, {Masana},
  {Messineo}, {Mowlavi}, {Nienartowicz}, {Panuzzo}, {Portell}, {Riello},
  {Seabroke}, {Tanga}, {Th{\'e}venin}, {Gracia-Abril}, {Comoretto},
  {Garcia-Reinaldos}, {Teyssier}, {Altmann}, {Andrae}, {Audard},
  {Bellas-Velidis}, {Benson}, {Berthier}, {Blomme}, {Burgess}, {Busso},
  {Carry}, {Cellino}, {Clementini}, {Clotet}, {Creevey}, {Davidson}, {De
  Ridder}, {Delchambre}, {Dell'Oro}, {Ducourant},
  {Fern{\'a}ndez-Hern{\'a}ndez}, {Fouesneau}, {Fr{\'e}mat}, {Galluccio},
  {Garc{\'\i}a-Torres}, {Gonz{\'a}lez-N{\'u}{\~n}ez}, {Gonz{\'a}lez-Vidal},
  {Gosset}, {Guy}, {Halbwachs}, {Hambly}, {Harrison}, {Hern{\'a}ndez},
  {Hestroffer}, {Hodgkin}, {Hutton}, {Jasniewicz}, {Jean-Antoine-Piccolo},
  {Jordan}, {Korn}, {Krone-Martins}, {Lanzafame}, {Lebzelter}, {L{\"o}ffler},
  {Manteiga}, {Marrese}, {Mart{\'\i}n-Fleitas}, {Moitinho}, {Mora}, {Muinonen},
  {Osinde}, {Pancino}, {Pauwels}, {Petit}, {Recio-Blanco}, {Richards},
  {Rimoldini}, {Robin}, {Sarro}, {Siopis}, {Smith}, {Sozzetti}, {S{\"u}veges},
  {Torra}, {van Reeven}, {Abbas}, {Abreu Aramburu}, {Accart}, {Aerts},
  {Altavilla}, {{\'A}lvarez}, {Alvarez}, {Alves}, {Anderson}, {Andrei},
  {Anglada Varela}, {Antiche}, {Antoja}, {Arcay}, {Astraatmadja}, {Bach},
  {Baker}, {Balaguer-N{\'u}{\~n}ez}, {Balm}, {Barache}, {Barata}, {Barbato},
  {Barblan}, {Barklem}, {Barrado}, {Barros}, {Barstow}, {Bartholom{\'e}
  Mu{\~n}oz}, {Bassilana}, {Becciani}, {Bellazzini}, {Berihuete}, {Bertone},
  {Bianchi}, {Bienaym{\'e}}, {Blanco-Cuaresma}, {Boch}, {Boeche}, {Bombrun},
  {Borrachero}, {Bossini}, {Bouquillon}, {Bourda}, {Bragaglia}, {Bramante},
  {Breddels}, {Bressan}, {Brouillet}, {Br{\"u}semeister}, {Brugaletta},
  {Bucciarelli}, {Burlacu}, {Busonero}, {Butkevich}, {Buzzi}, {Caffau},
  {Cancelliere}, {Cannizzaro}, {Cantat-Gaudin}, {Carballo}, {Carlucci},
  {Carrasco}, {Casamiquela}, {Castellani}, {Castro-Ginard}, {Charlot},
  {Chemin}, {Chiavassa}, {Cocozza}, {Costigan}, {Cowell}, {Crifo}, {Crosta},
  {Crowley}, {Cuypers}, {Dafonte}, {Damerdji}, {Dapergolas}, {David}, {David},
  {de Laverny}, {De Luise}, {De March}, {de Martino}, {de Souza}, {de Torres},
  {Debosscher}, {del Pozo}, {Delbo}, {Delgado}, {Delgado}, {Di Matteo},
  {Diakite}, {Diener}, {Distefano}, {Dolding}, {Drazinos}, {Dur{\'a}n},
  {Edvardsson}, {Enke}, {Eriksson}, {Esquej}, {Eynard Bontemps}, {Fabre},
  {Fabrizio}, {Faigler}, {Falc{\~a}o}, {Farr{\`a}s Casas}, {Federici},
  {Fedorets}, {Fernique}, {Figueras}, {Filippi}, {Findeisen}, {Fonti},
  {Fraile}, {Fraser}, {Fr{\'e}zouls}, {Gai}, {Galleti}, {Garabato},
  {Garc{\'\i}a-Sedano}, {Garofalo}, {Garralda}, {Gavel}, {Gavras}, {Gerssen},
  {Geyer}, {Giacobbe}, {Gilmore}, {Girona}, {Giuffrida}, {Glass}, {Gomes},
  {Granvik}, {Gueguen}, {Guerrier}, {Guiraud}, {Guti{\'e}rrez-S{\'a}nchez},
  {Haigron}, {Hatzidimitriou}, {Hauser}, {Haywood}, {Heiter}, {Helmi}, {Heu},
  {Hilger}, {Hobbs}, {Hofmann}, {Holland}, {Huckle}, {Hypki}, {Icardi},
  {Jan{\ss}en}, {Jevardat de Fombelle}, {Jonker}, {Juh{\'a}sz}, {Julbe},
  {Karampelas}, {Kewley}, {Klar}, {Kochoska}, {Kohley}, {Kolenberg},
  {Kontizas}, {Kontizas}, {Koposov}, {Kordopatis}, {Kostrzewa-Rutkowska},
  {Koubsky}, {Lambert}, {Lanza}, {Lasne}, {Lavigne}, {Le Fustec}, {Le
  Poncin-Lafitte}, {Lebreton}, {Leccia}, {Leclerc}, {Lecoeur-Taibi},
  {Lenhardt}, {Leroux}, {Liao}, {Licata}, {Lindstr{\o}m}, {Lister}, {Livanou},
  {Lobel}, {L{\'o}pez}, {Managau}, {Mann}, {Mantelet}, {Marchal}, {Marchant},
  {Marconi}, {Marinoni}, {Marschalk{\'o}}, {Marshall}, {Martino}, {Marton},
  {Mary}, {Massari}, {Matijevi{\v{c}}}, {Mazeh}, {McMillan}, {Messina},
  {Michalik}, {Millar}, {Molina}, {Molinaro}, {Moln{\'a}r}, {Montegriffo},
  {Mor}, {Morbidelli}, {Morel}, {Morris}, {Mulone}, {Muraveva}, {Musella},
  {Nelemans}, {Nicastro}, {Noval}, {O'Mullane}, {Ord{\'e}novic},
  {Ord{\'o}{\~n}ez-Blanco}, {Osborne}, {Pagani}, {Pagano}, {Pailler},
  {Palacin}, {Palaversa}, {Panahi}, {Pawlak}, {Piersimoni}, {Pineau}, {Plachy},
  {Plum}, {Poggio}, {Poujoulet}, {Pr{\v{s}}a}, {Pulone}, {Racero}, {Ragaini},
  {Rambaux}, {Ramos-Lerate}, {Regibo}, {Reyl{\'e}}, {Riclet}, {Ripepi}, {Riva},
  {Rivard}, {Rixon}, {Roegiers}, {Roelens}, {Romero-G{\'o}mez}, {Rowell},
  {Royer}, {Ruiz-Dern}, {Sadowski}, {Sagrist{\`a} Sell{\'e}s}, {Sahlmann},
  {Salgado}, {Salguero}, {Sanna}, {Santana-Ros}, {Sarasso}, {Savietto},
  {Schultheis}, {Sciacca}, {Segol}, {Segovia}, {S{\'e}gransan}, {Shih},
  {Siltala}, {Silva}, {Smart}, {Smith}, {Solano}, {Solitro}, {Sordo}, {Soria
  Nieto}, {Souchay}, {Spagna}, {Spoto}, {Stampa}, {Steele},
  {Steidelm{\"u}ller}, {Stephenson}, {Stoev}, {Suess}, {Surdej}, {Szabados},
  {Szegedi-Elek}, {Tapiador}, {Taris}, {Tauran}, {Taylor}, {Teixeira},
  {Terrett}, {Teyssand ier}, {Thuillot}, {Titarenko}, {Torra Clotet}, {Turon},
  {Ulla}, {Utrilla}, {Uzzi}, {Vaillant}, {Valentini}, {Valette}, {van Elteren},
  {Van Hemelryck}, {van Leeuwen}, {Vaschetto}, {Vecchiato}, {Veljanoski},
  {Viala}, {Vicente}, {Vogt}, {von Essen}, {Voss}, {Votruba}, {Voutsinas},
  {Walmsley}, {Weiler}, {Wertz}, {Wevers}, {Wyrzykowski}, {Yoldas},
  {{\v{Z}}erjal}, {Ziaeepour}, {Zorec}, {Zschocke}, {Zucker}, {Zurbach}, \&
  {Zwitter}}]{gai18}
{Gaia Collaboration}, {Brown}, A.~G.~A., {Vallenari}, A., {et~al.} 2018, \aap,
  616, A1

\bibitem[{{Geier} {et~al.}(2007){Geier}, {Nesslinger}, {Heber}, {Przybilla},
  {Napiwotzki}, \& {Kudritzki}}]{gei07}
{Geier}, S., {Nesslinger}, S., {Heber}, U., {et~al.} 2007, \aap, 464, 299

\bibitem[{{Geier} {et~al.}(2019){Geier}, {Raddi}, {Gentile Fusillo}, \&
  {Marsh}}]{gei19}
{Geier}, S., {Raddi}, R., {Gentile Fusillo}, N.~P., \& {Marsh}, T.~R. 2019,
  \aap, 621, A38

\bibitem[{{Geier} {et~al.}(2011){Geier}, {Hirsch}, {Tillich}, {Maxted},
  {Bentley}, {{\O}stensen}, {Heber}, {G{\"a}nsicke}, {Marsh}, {Napiwotzki},
  {Barlow}, \& {O'Toole}}]{gei11a}
{Geier}, S., {Hirsch}, H., {Tillich}, A., {et~al.} 2011, A\&A, 530, A28

\bibitem[{{Geier} {et~al.}(2013){Geier}, {Marsh}, {Wang}, {Dunlap}, {Barlow},
  {Schaffenroth}, {Chen}, {Irrgang}, {Maxted}, {Ziegerer}, {Kupfer},
  {Miszalski}, {Heber}, {Han}, {Shporer}, {Telting}, {G{\"a}nsicke},
  {{\O}stensen}, {O'Toole}, \& {Napiwotzki}}]{gei13}
{Geier}, S., {Marsh}, T.~R., {Wang}, B., {et~al.} 2013, \aap, 554, A54

\bibitem[{{Geier} {et~al.}(2015){Geier}, {F{\"u}rst}, {Ziegerer}, {Kupfer},
  {Heber}, {Irrgang}, {Wang}, {Liu}, {Han}, {Sesar}, {Levitan}, {Kotak},
  {Magnier}, {Smith}, {Burgett}, {Chambers}, {Flewelling}, {Kaiser},
  {Wainscoat}, \& {Waters}}]{gei15}
{Geier}, S., {F{\"u}rst}, F., {Ziegerer}, E., {et~al.} 2015, Science, 347, 1126

\bibitem[{{Graham} {et~al.}(2019){Graham}, {Kulkarni}, {Bellm}, {Adams},
  {Barbarino}, {Blagorodnova}, {Bodewits}, {Bolin}, {Brady}, {Cenko}, {Chang},
  {Coughlin}, {De}, {Eadie}, {Farnham}, {Feindt}, {Franckowiak}, {Fremling},
  {Gezari}, {Ghosh}, {Goldstein}, {Golkhou}, {Goobar}, {Ho}, {Huppenkothen},
  {Ivezi{\'c}}, {Jones}, {Juric}, {Kaplan}, {Kasliwal}, {Kelley}, {Kupfer},
  {Lee}, {Lin}, {Lunnan}, {Mahabal}, {Miller}, {Ngeow}, {Nugent}, {Ofek},
  {Prince}, {Rauch}, {van Roestel}, {Schulze}, {Singer}, {Sollerman}, {Taddia},
  {Yan}, {Ye}, {Yu}, {Barlow}, {Bauer}, {Beck}, {Belicki}, {Biswas}, {Brinnel},
  {Brooke}, {Bue}, {Bulla}, {Burruss}, {Connolly}, {Cromer}, {Cunningham},
  {Dekany}, {Delacroix}, {Desai}, {Duev}, {Feeney}, {Flynn}, {Frederick},
  {Gal-Yam}, {Giomi}, {Groom}, {Hacopians}, {Hale}, {Helou}, {Henning},
  {Hover}, {Hillenbrand}, {Howell}, {Hung}, {Imel}, {Ip}, {Jackson}, {Kaspi},
  {Kaye}, {Kowalski}, {Kramer}, {Kuhn}, {Landry}, {Laher}, {Mao}, {Masci},
  {Monkewitz}, {Murphy}, {Nordin}, {Patterson}, {Penprase}, {Porter},
  {Rebbapragada}, {Reiley}, {Riddle}, {Rigault}, {Rodriguez}, {Rusholme}, {van
  Santen}, {Shupe}, {Smith}, {Soumagnac}, {Stein}, {Surace}, {Szkody}, {Terek},
  {Van Sistine}, {van Velzen}, {Vestrand}, {Walters}, {Ward}, {Zhang}, \&
  {Zolkower}}]{gra19}
{Graham}, M.~J., {Kulkarni}, S.~R., {Bellm}, E.~C., {et~al.} 2019, \pasp, 131,
  078001

\bibitem[{{Green} {et~al.}(2019){Green}, {Schlafly}, {Zucker}, {Speagle}, \&
  {Finkbeiner}}]{gre19}
{Green}, G.~M., {Schlafly}, E.~F., {Zucker}, C., {Speagle}, J.~S., \&
  {Finkbeiner}, D.~P. 2019, arXiv e-prints, arXiv:1905.02734

\bibitem[{{Han} {et~al.}(2003){Han}, {Podsiadlowski}, {Maxted}, \&
  {Marsh}}]{han03}
{Han}, Z., {Podsiadlowski}, P., {Maxted}, P.~F.~L., \& {Marsh}, T.~R. 2003,
  MNRAS, 341, 669

\bibitem[{{Han} {et~al.}(2002){Han}, {Podsiadlowski}, {Maxted}, {Marsh}, \&
  {Ivanova}}]{han02}
{Han}, Z., {Podsiadlowski}, P., {Maxted}, P.~F.~L., {Marsh}, T.~R., \&
  {Ivanova}, N. 2002, MNRAS, 336, 449

\bibitem[{{Heber}(1986)}]{heb86}
{Heber}, U. 1986, \aap, 155, 33

\bibitem[{{Heber}(2009)}]{heb09}
---. 2009, \araa, 47, 211

\bibitem[{{Heber}(2016)}]{heb16}
---. 2016, \pasp, 128, 082001

\bibitem[{Hunter(2007)}]{hun07}
Hunter, J.~D. 2007, Computing In Science \& Engineering, 9, 90

\bibitem[{{Iben} \& {Tutukov}(1991)}]{ibe91}
{Iben}, Jr., I., \& {Tutukov}, A.~V. 1991, ApJ, 370, 615

\bibitem[{{Irrgang} {et~al.}(2013){Irrgang}, {Wilcox}, {Tucker}, \&
  {Schiefelbein}}]{irr13}
{Irrgang}, A., {Wilcox}, B., {Tucker}, E., \& {Schiefelbein}, L. 2013, \aap,
  549, A137

\bibitem[{{Israel} {et~al.}(1995){Israel}, {Stella}, {Angelini}, {White}, \&
  {Giommi}}]{isr95}
{Israel}, G.~L., {Stella}, L., {Angelini}, L., {White}, N.~E., \& {Giommi}, P.
  1995, \iaucirc, 6277, 1

\bibitem[{{Israel} {et~al.}(1997){Israel}, {Stella}, {Angelini}, {White},
  {Kallman}, {Giommi}, \& {Treves}}]{isr97}
{Israel}, G.~L., {Stella}, L., {Angelini}, L., {et~al.} 1997, \apjl, 474, L53

\bibitem[{{Kalirai} {et~al.}(2014){Kalirai}, {Marigo}, \& {Tremblay}}]{kal14}
{Kalirai}, J.~S., {Marigo}, P., \& {Tremblay}, P.-E. 2014, \apj, 782, 17

\bibitem[{{King} \& {Ritter}(1998)}]{kin98}
{King}, A.~R., \& {Ritter}, H. 1998, \mnras, 293, L42

\bibitem[{{Koen} {et~al.}(1998){Koen}, {Orosz}, \& {Wade}}]{koe98}
{Koen}, C., {Orosz}, J.~A., \& {Wade}, R.~A. 1998, MNRAS, 300, 695

\bibitem[{{Kollmeier} {et~al.}(2017){Kollmeier}, {Zasowski}, {Rix}, {Johns},
  {Anderson}, {Drory}, {Johnson}, {Pogge}, {Bird}, {Blanc}, {Brownstein},
  {Crane}, {De Lee}, {Klaene}, {Kreckel}, {MacDonald}, {Merloni}, {Ness},
  {O'Brien}, {Sanchez-Gallego}, {Sayres}, {Shen}, {Thakar}, {Tkachenko},
  {Aerts}, {Blanton}, {Eisenstein}, {Holtzman}, {Maoz}, {Nandra}, {Rockosi},
  {Weinberg}, {Bovy}, {Casey}, {Chaname}, {Clerc}, {Conroy}, {Eracleous},
  {G{\"a}nsicke}, {Hekker}, {Horne}, {Kauffmann}, {McQuinn}, {Pellegrini},
  {Schinnerer}, {Schlafly}, {Schwope}, {Seibert}, {Teske}, \& {van
  Saders}}]{kol17}
{Kollmeier}, J.~A., {Zasowski}, G., {Rix}, H.-W., {et~al.} 2017, arXiv
  e-prints, arXiv:1711.03234

\bibitem[{{Kudritzki} \& {Simon}(1978)}]{kud78}
{Kudritzki}, R.~P., \& {Simon}, K.~P. 1978, \aap, 70, 653

\bibitem[{{Kupfer} {et~al.}(2015){Kupfer}, {Geier}, {Heber}, {{\O}stensen},
  {Barlow}, {Maxted}, {Heuser}, {Schaffenroth}, \& {G{\"a}nsicke}}]{kup15a}
{Kupfer}, T., {Geier}, S., {Heber}, U., {et~al.} 2015, A\&A, 576, A44

\bibitem[{{Kupfer} {et~al.}(2017{\natexlab{a}}){Kupfer}, {van Roestel},
  {Brooks}, {Geier}, {Marsh}, {Groot}, {Bloemen}, {Prince}, {Bellm}, {Heber},
  {Bildsten}, {Miller}, {Dyer}, {Dhillon}, {Green}, {Irawati}, {Laher},
  {Littlefair}, {Shupe}, {Steidel}, {Rattansoon}, \& {Pettini}}]{kup17}
{Kupfer}, T., {van Roestel}, J., {Brooks}, J., {et~al.} 2017{\natexlab{a}},
  \apj, 835, 131

\bibitem[{{Kupfer} {et~al.}(2017{\natexlab{b}}){Kupfer}, {Ramsay}, {van
  Roestel}, {Brooks}, {MacFarlane}, {Toma}, {Groot}, {Woudt}, {Bildsten},
  {Marsh}, {Green}, {Breedt}, {Kilkenny}, {Freudenthal}, {Geier}, {Heber},
  {Bagnulo}, {Blagorodnova}, {Buckley}, {Dhillon}, {Kulkarni}, {Lunnan}, \&
  {Prince}}]{kup17a}
{Kupfer}, T., {Ramsay}, G., {van Roestel}, J., {et~al.} 2017{\natexlab{b}},
  \apj, 851, 28

\bibitem[{{Kupfer} {et~al.}(2018){Kupfer}, {Korol}, {Shah}, {Nelemans},
  {Marsh}, {Ramsay}, {Groot}, {Steeghs}, \& {Rossi}}]{kup18}
{Kupfer}, T., {Korol}, V., {Shah}, S., {et~al.} 2018, \mnras, 480, 302

\bibitem[{{Livne}(1990)}]{liv90}
{Livne}, E. 1990, ApJl, 354, L53

\bibitem[{{Livne} \& {Arnett}(1995)}]{liv95}
{Livne}, E., \& {Arnett}, D. 1995, ApJ, 452, 62

\bibitem[{{Loeb} \& {Gaudi}(2003)}]{loe03}
{Loeb}, A., \& {Gaudi}, B.~S. 2003, \apjl, 588, L117

\bibitem[{{Macfarlane} {et~al.}(2015){Macfarlane}, {Toma}, {Ramsay}, {Groot},
  {Woudt}, {Drew}, {Barentsen}, \& {Eisl{\"o}ffel}}]{mac15}
{Macfarlane}, S.~A., {Toma}, R., {Ramsay}, G., {et~al.} 2015, MNRAS, 454, 507

\bibitem[{{Macfarlane} {et~al.}(2017){Macfarlane}, {Woudt}, {Groot}, {Ramsay},
  {Toma}, {Motsoaledi}, {Crause}, {Gilbank}, {O'Donoghue}, {Potter},
  {Sickafoose}, {van Gend}, \& {Worters}}]{mac17}
{Macfarlane}, S.~A., {Woudt}, P.~A., {Groot}, P.~J., {et~al.} 2017, MNRAS, 465,
  434

\bibitem[{{Marsh} {et~al.}(2004){Marsh}, {Nelemans}, \& {Steeghs}}]{mar04}
{Marsh}, T.~R., {Nelemans}, G., \& {Steeghs}, D. 2004, MNRAS, 350, 113

\bibitem[{{Masci} {et~al.}(2019){Masci}, {Laher}, {Rusholme}, {Shupe}, {Groom},
  {Surace}, {Jackson}, {Monkewitz}, {Beck}, {Flynn}, {Terek}, {Landry},
  {Hacopians}, {Desai}, {Howell}, {Brooke}, {Imel}, {Wachter}, {Ye}, {Lin},
  {Cenko}, {Cunningham}, {Rebbapragada}, {Bue}, {Miller}, {Mahabal}, {Bellm},
  {Patterson}, {Juri{\'c}}, {Golkhou}, {Ofek}, {Walters}, {Graham}, {Kasliwal},
  {Dekany}, {Kupfer}, {Burdge}, {Cannella}, {Barlow}, {Van Sistine}, {Giomi},
  {Fremling}, {Blagorodnova}, {Levitan}, {Riddle}, {Smith}, {Helou}, {Prince},
  \& {Kulkarni}}]{mas19}
{Masci}, F.~J., {Laher}, R.~R., {Rusholme}, B., {et~al.} 2019, \pasp, 131,
  018003

\bibitem[{{Maxted} {et~al.}(2001){Maxted}, {Heber}, {Marsh}, \&
  {North}}]{max01}
{Maxted}, P.~f.~L., {Heber}, U., {Marsh}, T.~R., \& {North}, R.~C. 2001, MNRAS,
  326, 1391

\bibitem[{{Maxted} {et~al.}(2000){Maxted}, {Marsh}, \& {North}}]{max00}
{Maxted}, P.~F.~L., {Marsh}, T.~R., \& {North}, R.~C. 2000, MNRAS, 317, L41

\bibitem[{{McCarthy} {et~al.}(1998){McCarthy}, {Cohen}, {Butcher}, {Cromer},
  {Croner}, {Douglas}, {Goeden}, {Grewal}, {Lu}, {Petrie}, {Weng}, {Weber},
  {Koch}, \& {Rodgers}}]{mcc98}
{McCarthy}, J.~K., {Cohen}, J.~G., {Butcher}, B., {et~al.} 1998, in \procspie,
  Vol. 3355, Optical Astronomical Instrumentation, ed. S.~{D'Odorico}, 81--92

\bibitem[{{Mereghetti} \& {La Palombara}(2016)}]{mer16}
{Mereghetti}, S., \& {La Palombara}, N. 2016, Advances in Space Research, 58,
  809

\bibitem[{{Mereghetti} {et~al.}(2011){Mereghetti}, {La Palombara}, {Tiengo},
  {Pizzolato}, {Esposito}, {Woudt}, {Israel}, \& {Stella}}]{mer11}
{Mereghetti}, S., {La Palombara}, N., {Tiengo}, A., {et~al.} 2011, \apj, 737,
  51

\bibitem[{{Mereghetti} {et~al.}(2013){Mereghetti}, {La Palombara}, {Tiengo},
  {Sartore}, {Esposito}, {Israel}, \& {Stella}}]{mer13}
---. 2013, \aap, 553, A46

\bibitem[{{Mereghetti} {et~al.}(2009){Mereghetti}, {Tiengo}, {Esposito}, {La
  Palombara}, {Israel}, \& {Stella}}]{mer09}
{Mereghetti}, S., {Tiengo}, A., {Esposito}, P., {et~al.} 2009, Science, 325,
  1222

\bibitem[{{Morris}(1985)}]{1985ApJ...295..143M}
{Morris}, S.~L. 1985, \apj, 295, 143

\bibitem[{{Napiwotzki} {et~al.}(2004){Napiwotzki}, {Karl}, {Lisker}, {Heber},
  {Christlieb}, {Reimers}, {Nelemans}, \& {Homeier}}]{nap04a}
{Napiwotzki}, R., {Karl}, C.~A., {Lisker}, T., {et~al.} 2004, Astrophysics and
  Space Science, 291, 321

\bibitem[{{Nelemans}(2010)}]{nel10a}
{Nelemans}, G. 2010, \apss, 329, 25

\bibitem[{{Nomoto}(1982)}]{Nomoto82}
{Nomoto}, K. 1982, \apj, 253, 798

\bibitem[{{Nomoto} {et~al.}(2007){Nomoto}, {Saio}, {Kato}, \&
  {Hachisu}}]{Nomoto07}
{Nomoto}, K., {Saio}, H., {Kato}, M., \& {Hachisu}, I. 2007, \apj, 663, 1269

\bibitem[{{Odenkirchen} \& {Brosche}(1992)}]{ode92}
{Odenkirchen}, M., \& {Brosche}, P. 1992, Astronomische Nachrichten, 313, 69

\bibitem[{{Oke} \& {Gunn}(1982)}]{oke82}
{Oke}, J.~B., \& {Gunn}, J.~E. 1982, PASP, 94, 586

\bibitem[{Oliphant(2015)}]{numpy}
Oliphant, T.~E. 2015, Guide to NumPy, 2nd edn. (USA: CreateSpace Independent
  Publishing Platform)

\bibitem[{{Orosz} \& {Wade}(1999)}]{oro99}
{Orosz}, J.~A., \& {Wade}, R.~A. 1999, MNRAS, 310, 773

\bibitem[{{Patterson} \& {Raymond}(1985)}]{pat85}
{Patterson}, J., \& {Raymond}, J.~C. 1985, \apj, 292, 535

\bibitem[{{Pauli} {et~al.}(2006){Pauli}, {Napiwotzki}, {Heber}, {Altmann}, \&
  {Odenkirchen}}]{pau06}
{Pauli}, E.-M., {Napiwotzki}, R., {Heber}, U., {Altmann}, M., \& {Odenkirchen},
  M. 2006, \aap, 447, 173

\bibitem[{{Paxton} {et~al.}(2011){Paxton}, {Bildsten}, {Dotter}, {Herwig},
  {Lesaffre}, \& {Timmes}}]{pax11}
{Paxton}, B., {Bildsten}, L., {Dotter}, A., {et~al.} 2011, ApJs, 192, 3

\bibitem[{{Paxton} {et~al.}(2013){Paxton}, {Cantiello}, {Arras}, {Bildsten},
  {Brown}, {Dotter}, {Mankovich}, {Montgomery}, {Stello}, {Timmes}, \&
  {Townsend}}]{pax13}
{Paxton}, B., {Cantiello}, M., {Arras}, P., {et~al.} 2013, ApJs, 208, 4

\bibitem[{{Paxton} {et~al.}(2015){Paxton}, {Marchant}, {Schwab}, {Bauer},
  {Bildsten}, {Cantiello}, {Dessart}, {Farmer}, {Hu}, {Langer}, {Townsend},
  {Townsley}, \& {Timmes}}]{pax15}
{Paxton}, B., {Marchant}, P., {Schwab}, J., {et~al.} 2015, ApJs, 220, 15

\bibitem[{{Paxton} {et~al.}(2018){Paxton}, {Schwab}, {Bauer}, {Bildsten},
  {Blinnikov}, {Duffell}, {Farmer}, {Goldberg}, {Marchant}, {Sorokina},
  {Thoul}, {Townsend}, \& {Timmes}}]{pax18}
{Paxton}, B., {Schwab}, J., {Bauer}, E.~B., {et~al.} 2018, \apjs, 234, 34

\bibitem[{{Paxton} {et~al.}(2019){Paxton}, {Smolec}, {Schwab}, {Gautschy},
  {Bildsten}, {Cantiello}, {Dotter}, {Farmer}, {Goldberg}, {Jermyn}, {Kanbur},
  {Marchant}, {Thoul}, {Townsend}, {Wolf}, {Zhang}, \& {Timmes}}]{pax19}
{Paxton}, B., {Smolec}, R., {Schwab}, J., {et~al.} 2019, \apjs, 243, 10

\bibitem[{{Perets} {et~al.}(2019){Perets}, {Zenati}, {Toonen}, \&
  {Bobrick}}]{Perets19}
{Perets}, H.~B., {Zenati}, Y., {Toonen}, S., \& {Bobrick}, A. 2019, arXiv
  e-prints, arXiv:1910.07532

\bibitem[{{Perley}(2019)}]{per19}
{Perley}, D.~A. 2019, arXiv e-prints, arXiv:1903.07629

\bibitem[{{Piersanti} {et~al.}(2014){Piersanti}, {Tornamb{\'e}}, \&
  {Yungelson}}]{pie14}
{Piersanti}, L., {Tornamb{\'e}}, A., \& {Yungelson}, L.~R. 2014, MNRAS, 445,
  3239

\bibitem[{{Polin} {et~al.}(2019){Polin}, {Nugent}, \& {Kasen}}]{pol19}
{Polin}, A., {Nugent}, P., \& {Kasen}, D. 2019, \apj, 873, 84

\bibitem[{{Pringle} \& {Savonije}(1979)}]{pri79}
{Pringle}, J.~E., \& {Savonije}, G.~J. 1979, \mnras, 187, 777

\bibitem[{{Ramsay} {et~al.}(2019){Ramsay}, {Kennedy}, {Hakala}, \&
  {Jeffery}}]{ram19}
{Ramsay}, G., {Kennedy}, M., {Hakala}, P., \& {Jeffery}, C.~S. 2019, The
  Astronomer's Telegram, 13048, 1

\bibitem[{{Ratzloff} {et~al.}(2019){Ratzloff}, {Barlow}, {Kupfer}, {Corcoran},
  {Geier}, {Bauer}, {Corbett}, {Howard}, {Glazier}, \& {Law}}]{rat19}
{Ratzloff}, J.~K., {Barlow}, B.~N., {Kupfer}, T., {et~al.} 2019, \apj, 883, 51

\bibitem[{{Ritter}(1988)}]{Ritter1988}
{Ritter}, H. 1988, \aap, 202, 93

\bibitem[{{Rivera Sandoval} {et~al.}(2019){Rivera Sandoval}, {Maccarone}, \&
  {Murawski}}]{riv19}
{Rivera Sandoval}, L.~E., {Maccarone}, T., \& {Murawski}, G. 2019, The
  Astronomer's Telegram, 12847, 1

\bibitem[{{Saio}(2008)}]{sai08}
{Saio}, H. 2008, in Astronomical Society of the Pacific Conference Series, Vol.
  391, Hydrogen-Deficient Stars, ed. A.~{Werner} \& T.~{Rauch}, 69

\bibitem[{{Savonije} {et~al.}(1986){Savonije}, {de Kool}, \& {van den
  Heuvel}}]{sav86}
{Savonije}, G.~J., {de Kool}, M., \& {van den Heuvel}, E.~P.~J. 1986, A\&A,
  155, 51

\bibitem[{{Scargle}(1982)}]{sca82}
{Scargle}, J.~D. 1982, \apj, 263, 835

\bibitem[{{Shen} \& {Bildsten}(2014)}]{she14}
{Shen}, K.~J., \& {Bildsten}, L. 2014, ApJ, 785, 61

\bibitem[{{Stroeer} {et~al.}(2007){Stroeer}, {Heber}, {Lisker}, {Napiwotzki},
  {Dreizler}, {Christlieb}, \& {Reimers}}]{str07}
{Stroeer}, A., {Heber}, U., {Lisker}, T., {et~al.} 2007, A\&A, 462, 269

\bibitem[{{Thackeray}(1970)}]{tha70}
{Thackeray}, A.~D. 1970, \mnras, 150, 215

\bibitem[{{Toma} {et~al.}(2016){Toma}, {Ramsay}, {Macfarlane}, {Groot},
  {Woudt}, {Dhillon}, {Jeffery}, {Marsh}, {Nelemans}, \& {Steeghs}}]{tom16}
{Toma}, R., {Ramsay}, G., {Macfarlane}, S., {et~al.} 2016, MNRAS, 463, 1099

\bibitem[{{Tutukov} \& {Fedorova}(1989)}]{tut89}
{Tutukov}, A.~V., \& {Fedorova}, A.~V. 1989, Soviet Astronomy, 33, 606

\bibitem[{{Tutukov} \& {Yungelson}(1990)}]{tut90}
{Tutukov}, A.~V., \& {Yungelson}, L.~R. 1990, \sovast, 34, 57

\bibitem[{Vanderplas(2015)}]{van15a}
Vanderplas, J. 2015, {gatspy: General tools for Astronomical Time Series in
  Python}, vv0.3.0,  Zenodo, doi:10.5281/zenodo.14833.
\newblock \url{https://doi.org/10.5281/zenodo.14833}

\bibitem[{{VanderPlas} \& {Ivezi\'{c}}(2015)}]{van15}
{VanderPlas}, J.~T., \& {Ivezi\'{c}}, v. 2015, \apj, 812, 18

\bibitem[{{Vennes} {et~al.}(2012){Vennes}, {Kawka}, {O'Toole}, {N{\'e}meth}, \&
  {Burton}}]{ven12}
{Vennes}, S., {Kawka}, A., {O'Toole}, S.~J., {N{\'e}meth}, P., \& {Burton}, D.
  2012, \apjl, 759, L25

\bibitem[{{Wang}(2018)}]{wan18}
{Wang}, B. 2018, Research in Astronomy and Astrophysics, 18, 049

\bibitem[{{Wang} \& {Han}(2012)}]{wan12}
{Wang}, B., \& {Han}, Z. 2012, \nar, 56, 122

\bibitem[{{Wheatley} {et~al.}(2003){Wheatley}, {Mauche}, \& {Mattei}}]{whe03}
{Wheatley}, P.~J., {Mauche}, C.~W., \& {Mattei}, J.~A. 2003, \mnras, 345, 49

\bibitem[{{Wolf} {et~al.}(2013){Wolf}, {Bildsten}, {Brooks}, \&
  {Paxton}}]{Wolf13}
{Wolf}, W.~M., {Bildsten}, L., {Brooks}, J., \& {Paxton}, B. 2013, \apj, 777,
  136

\bibitem[{{Woosley} \& {Kasen}(2011)}]{woo11}
{Woosley}, S.~E., \& {Kasen}, D. 2011, ApJ, 734, 38

\bibitem[{{Yungelson}(2008)}]{yun08}
{Yungelson}, L.~R. 2008, Astronomy Letters, 34, 620

\bibitem[{{Zenati} {et~al.}(2019){Zenati}, {Toonen}, \& {Perets}}]{Zenati19}
{Zenati}, Y., {Toonen}, S., \& {Perets}, H.~B. 2019, \mnras, 482, 1135

\end{thebibliography}

\end{document}